\documentclass[twocolumn,showkeys,showpacs,preprintnumbers,prd,superscriptaddress,nofootinbib,aps,10pt]{revtex4-1}
\usepackage{graphicx,epsf,bm,amsmath,amsfonts,amssymb,epstopdf,natbib,color,verbatim,multirow,bm,mathtools,mathrsfs,braket,bbold,xcolor}
\usepackage{hyperref}
\usepackage[normalem]{ulem}
\hypersetup{colorlinks=true,urlcolor=blue,citecolor=blue,linkcolor=blue,menucolor=blue,anchorcolor=blue,filecolor=blue}
\date{\today}

\begin{document}

\title{Direct detection of solar chameleons with electron recoil data from XENONnT}

\author{Guan-Wen Yuan}
\email{guanwen.yuan@unitn.it}
\affiliation{Department of Physics, University of Trento, Via Sommarive 14, 38123 Povo (TN), Italy \looseness=-1}
\affiliation{Trento Institute for Fundamental Physics and Applications (TIFPA)-INFN, Via Sommarive 14, 38123 Povo (TN), Italy \looseness=-1}

\author{Anne-Christine Davis}
\email{ad107@cam.ac.uk}
\affiliation{Department of Applied Mathematics and Theoretical Physics (DAMTP), University of Cambridge, CB3 0WA, United Kingdom \looseness=-1}
\affiliation{Kavli Institute for Cosmology (KICC), University of Cambridge, Madingley Road, Cambridge CB3 0HA, United Kingdom \looseness=-1}

\author{Maurizio Giannotti}
\email{mgiannotti@unizar.es}
\affiliation{Departamento de F{\'i}sica Te{\'o}rica, Universidad de Zaragoza, C.\ de Pedro Cerbuna 12, 50009 Zaragoza, Spain \looseness=-1}
\affiliation{Centro de Astropartículas y F{\'i}sica de Altas Energ{\'i}as, Universidad de Zaragoza, C.\ de Pedro Cerbuna 12, 50009 Zaragoza, Spain \looseness=-1}

\author{Sunny Vagnozzi}
\email{sunny.vagnozzi@unitn.it}
\affiliation{Department of Physics, University of Trento, Via Sommarive 14, 38123 Povo (TN), Italy \looseness=-1}
\affiliation{Trento Institute for Fundamental Physics and Applications (TIFPA)-INFN, Via Sommarive 14, 38123 Povo (TN), Italy \looseness=-1}

\author{Luca Visinelli}
\email{lvisinelli@unisa.it}
\affiliation{Department of Physics, University of Salerno, Via Giovanni Paolo II 132, 84084 Fisciano (SA), Italy \looseness=-1}
\affiliation{INFN Sezione di Napoli, Gruppo Collegato di Salerno, Via Giovanni Paolo II 132, 84084 Fisciano (SA), Italy \looseness=-1}

\author{Julia K. Vogel}
\email{julia.vogel@tu-dortmund.de}
\affiliation{Fakult\"{a}t f\"{u}r Physik, Technische Universit\"{a}t Dortmund, Otto-Hahn-Stra{\ss}e 4, 44221 Dortmund, Germany \looseness=-1}

\begin{abstract}
\noindent We reassess prospects for direct detection of solar chameleons, in light of recent progress in modeling their production, and the availability of new XENONnT data. We show that the contribution from Primakoff production in the electric fields of electrons and ions dominates the electron recoil event rate, which is enhanced compared to earlier estimates based on magnetic conversion in the tachocline alone. We argue that the signal is governed by the effective coupling $\beta_{\text{eff}} \equiv \beta_{\gamma}M_e^{-4}$, which encodes the combined effects of production and detection, where $\beta_{\gamma}$ and $M_e$ are the chameleon-photon (conformal) coupling and chameleon-electron disformal coupling scale, respectively. Setting the height of the chameleon potential to the dark energy (DE) scale $\Lambda \simeq 2.4\,{\text{meV}}$, we show that XENONnT electron recoil data set the upper limit $\log_{10}\beta_{\text{eff}}<-6.9$. This limit is independent of the conformal matter coupling $\beta_m$ and index $n$, and applies to the whole class of inverse power-law chameleons, well beyond the $n=1$ case usually studied. We comment on how future multi-target experiments and lower-threshold analyses could distinguish solar chameleons from other light (pseudo)scalar particles such as axions. Our work demonstrates that existing dark matter direct detection experiments can probe regions of parameter space relevant to screened DE models, providing complementary tests to astrophysical and fifth-force searches at no additional experimental cost.
\end{abstract}

\maketitle

\section{Introduction}
\label{sec:introduction}

New (ultra)light scalar fields are among the most actively explored extensions of the Standard Model (SM), motivating a broad program of experimental and theoretical studies~\cite{Arias:2012az, Marsh:2015xka, Irastorza:2018dyq, DiLuzio:2020wdo, Antypas:2022asj, Arza:2026rsl}. On the experimental side, several detection strategies for these particles are being pursued, including terrestrial experiments, precision atomic and neutron measurements, and probes of their cosmological and astrophysical imprints~\cite{Riotto:2000kh,Torres:2000dw,Adelberger:2003zx,Kapner:2006si,Serebrov:2009zz,Chiow:2011zz,Brito:2015oca,Hees:2016gop,Safronova:2017xyt,SimonsObservatory:2018koc,SimonsObservatory:2019qwx,Roy:2021uye,Odintsov:2022cbm,Yuan:2022nmu,Chen:2022nbb,Poulin:2023lkg,Kading:2023mdk,Alesini:2023qed,Kading:2024jqe,Calza:2025yfm}. From the theoretical perspective, light scalar fields arise naturally in a variety of beyond the SM constructions, including string compactifications~\cite{Svrcek:2006yi,Arvanitaki:2009fg,Visinelli:2018utg,Cicoli:2023opf,Gendler:2023kjt} and modified gravity theories~\cite{Sotiriou:2008rp,DeFelice:2010aj,Nojiri:2010wj,Clifton:2011jh,Chamseddine:2013kea,deRham:2014zqa,Cai:2015emx,Nojiri:2017ncd,Oikonomou:2020qah,Oikonomou:2022wuk,Hell:2023mph,Hell:2025wha,Heisenberg:2025fxc,Hell:2025lbl,Barker:2025gon}. Moreover, these particles may hold clues to the origin of dark matter (DM)~\cite{Hu:2000ke,Boehm:2003hm,Cline:2013gha,Foot:2014uba,Hui:2016ltb,Odintsov:2019mlf,Odintsov:2019evb,Ferreira:2020fam,Oikonomou:2023kqi,Oikonomou:2024geq}, as well as the physics underlying cosmic acceleration~\cite{Ratra:1987rm,Wetterich:1987fm,Caldwell:1997ii,Zlatev:1998tr,Copeland:2006wr,Tsujikawa:2013fta,Odintsov:2020nwm,Giare:2024sdl,Kaneta:2025kcn,Lin:2025gne}.

To account for cosmic acceleration, and therefore play a role as a viable dark energy (DE) candidate, a new scalar field needs to be extremely light, at least on cosmological scales. This extreme lightness immediately leads to a major obstacle: a canonically coupled ultralight scalar generically mediates a long-range ``fifth force'' between macroscopic bodies~\cite{Carroll:1998zi,Amendola:1999er}, in obvious conflict with laboratory and solar system tests of gravity~\cite{Adelberger:2003zx,Kapner:2006si,Will:2014kxa,Tsai:2021irw,Tsai:2023zza}.~\footnote{Pseudo-scalar (axion-like) particles evade fifth-force constraints because of their derivative couplings to matter. At leading order, these interactions do not mediate spin-independent forces between unpolarized macroscopic bodies~\cite{Moody:1984ba}.} Barring an extreme degree of fine-tuning, such forces must be hidden via a dynamical mechanism which suppresses their effects in high-density environments, while allowing the scalar to remain active on cosmological scales. This is the purpose of screening mechanisms, which come in several forms, including the chameleon~\cite{Khoury:2003aq,Khoury:2003rn}, symmetron~\cite{Hinterbichler:2010es,Brax:2011pk,Burrage:2018zuj}, environmental dilaton~\cite{Damour:1994zq,Brax:2010gi,Burrage:2025grx}, and Vainshtein mechanisms~\cite{Vainshtein:1972sx,Deffayet:2001uk}, each relying on distinct dynamics (see Refs.~\cite{Brax:2021wcv, Brax:2026cmh} for reviews). Depending on the choice of frame, these mechanisms can typically (as in the chameleon case) be viewed either as a new scalar non-minimally coupled to matter in the Einstein frame, or as scalar–tensor modifications of gravity in the Jordan frame, where the term ``screened modified gravity'' is often used~\cite{Sakstein:2014jrq,Joyce:2016vqv}.

Among the various screening mechanisms proposed, chameleon screening is one of the best studied examples (see e.g.\ Refs.~\cite{Brax:2004ym,Brax:2004qh,Capozziello:2007eu,Brax:2008hh,Brax:2010kv,Gannouji:2010fc,Wang:2012kj,Erickcek:2013dea,Elder:2016yxm,Brax:2016did,Burrage:2017shh,Burrage:2018pyg,Katsuragawa:2019uto,Sakstein:2019qgn,Desmond:2019ygn,Hartley:2019wzu,Cai:2021wgv,Karwal:2021vpk,Katsuragawa:2021wmw,Dima:2021pwx,Benisty:2021cmq,Tamosiunas:2021kth,Briddon:2021etm,Brax:2021owd,Yuan:2022cpw,Chakrabarti:2022zvv,Tamosiunas:2022tic,Brax:2022olf,Benisty:2022lox,Boumechta:2023qhd,Elder:2023oar,Benisty:2023dkn,Paliathanasis:2023ttu,Paliathanasis:2023dfz,Benisty:2023vbz,Briddon:2023ayq,Hogas:2023pjz,Zaregonbadi:2023vcv,Benisty:2023clf,Fischer:2023eww,Kumar:2024ylj,Fischer:2024coj,Baez-Camargo:2024jia,Fischer:2024eic,Paliathanasis:2024sle,Fischer:2024gni,Pizzuti:2024hym,Nojiri:2025low,Zaregonbadi:2025ils,Feleppa:2025clx,Neckam:2025kip,Feleppa:2025vop}). Through a direct coupling to the ambient matter density, chameleon-screened scalars (hereafter simply ``chameleons'') acquire a density-dependent effective mass $m_{\text{eff}}$, which increases in high-density environments. As a result, in such environments the fifth force they mediate becomes short-ranged and evades detection, while in low-density regions the field remains (ultra)light, allowing it to play a cosmological role as DE. At the same time, this coupling inevitably makes chameleons amenable to local (terrestrial) tests of their effects~\cite{Burrage:2016bwy,Burrage:2017qrf}.

A wide range of experimental approaches has been proposed in this sense, including but not limited to precision atomic and neutron spectroscopy, torsion balance experiments, terrestrial laboratory tests, entanglement-based setups and, last but not least, astrophysical and cosmological observations~\cite{Burrage:2016bwy,Burrage:2017qrf}. In addition to these detection strategies, a powerful astrophysical source of chameleons is provided by the Sun, where chameleons can be produced from a variety of mechanisms before eventually escaping, if sufficiently weakly coupled, and reaching the Earth~\cite{Brax:2010xq,Brax:2011wp}. The characteristic energies of solar chameleons are set by the thermal scales in the solar plasma, and are therefore in the ${\text{keV}}$ range, which is precisely the regime of recoil energies of interest to underground multi-tonne DM direct detection experiments. In 2021, some of us pointed out that this potentially allows for searches of chameleon particles in DM direct detection experiments~\cite{Vagnozzi:2021quy}. Motivated by the electron recoil excess reported at the time by the XENON1T experiment~\cite{XENON:2020rca}, a study using XENON1T electron recoil data was then carried out as a proof of principle for the feasibility of terrestrial direct detection of screened DE in Ref.~\cite{Vagnozzi:2021quy} (hereafter Paper~I).~\footnote{While the claim for an excess in the electron recoil spectrum was later ruled out~\cite{XENON:2022ltv}, the quest for scalar fields produced in the Sun has since flourished. The search is undergoing in other experiments, including xenon-based ones such as LUX-ZEPLIN~\cite{LZ:2023poo, LZ:2025igz} and PandaX~\cite{PandaX:2024cic}, germanium-based ones such as EDELWEISS~\cite{Armengaud:2013rta}, CDMS~\cite{CDMS:2009fba}, and CDEX-1B~\cite{Yang:2024lpa}, and argon-based ones such as DarkSide~\cite{DarkSide:2022knj}. Searches for (pseudo)scalar particles produced in the Sun are also being carried out via helioscopes such as the CERN Axion Solar Telescope (CAST)~\cite{CAST:2015npk, CAST:2018bce}, as well as the upcoming International Axion Observatory (IAXO) and BabyIAXO~\cite{IAXO:2019mpb, IAXO:2025ltd}. The search can be extended in the X-ray band, by considering conversion of scalars into photons in the Sun’s atmospheric magnetic field, and using X-ray data from the Nuclear Spectroscopic Telescope Array (NuSTAR)~\cite{Ruz:2024gkl}.}

Since this earlier proof-of-principle work, a number of theoretical and experimental developments motivate a fresh look at this scenario. On the theoretical side, significant refinements have been recently worked out by some of us for what concerns the calculation of the solar chameleon flux. More specifically, Ref.~\cite{OShea:2024jjw} (hereafter Paper~II) presents an improved treatment of chameleon production in the solar bulk magnetic field (earlier limited to a thin shell around the tachocline), while also including Primakoff production in the electric fields of electrons and ions: the result is a much more complete and reliable model for the flux of solar chameleons. On the experimental side, the XENON collaboration has progressed from XENON1T to XENONnT~\cite{XENON:2024wpa}, with a larger exposure and improved sensitivity, and no indications of the excess present earlier in XENON1T~\cite{XENON:2022ltv}. All of these considerations make it especially timely to revisit prospects for terrestrial detection of solar chameleons, adopting up-to-date, state-of-the-art solar flux calculations, and using the latest available data. In what follows, we shall remain agnostic for what concerns the connection of chameleons to cosmic acceleration. Instead, we will treat them simply as a theoretically consistent class of screened scalar particles.~\footnote{A well-known no-go theorem precludes the possibility of self-acceleration driven by the chameleon mechanism~\cite{Wang:2012kj}, but does not exclude the possibility of a chameleon field driving cosmic acceleration \textit{\`{a} la} quintessence, i.e.\ rolling down the potential rather than by means of modified gravity effects.}

The rest of this work is then organized as follows. In Sec.~\ref{sec:solarchameleons} we briefly review the physics of chameleons, as well as their production in the Sun. We then discuss our methodology to calculate the theoretical expectations for the event rate and place limits on the chameleon parameters from XENONnT electron recoil data in Sec.~\ref{sec:methodology}. Our results are presented and critically discussed in Sec.~\ref{sec:resultsdiscussion}. Finally, in Sec.~\ref{sec:conclusions} we draw closing remarks. The model we adopt for the radial profile of the solar magnetic field is discussed in Appendix~\ref{app:solarmodel}, whereas the full posterior distributions for the chameleon parameter resulting from an exploratory Bayesian analysis of XENONnT data are shown in Appendix~\ref{app:posteriors}. The code used for our analysis is publicly available on {\includegraphics[height=1.9ex]{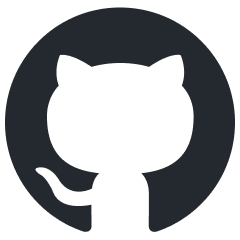}\,\texttt{GitHub} at \href{https://github.com/yuanguanwen/Chameleon\_Detection}{\texttt{github.com/yuanguanwen/Chameleon\_Detection}}. In what follows, we adopt natural units with $\hbar=c=k_B=1$, unless otherwise stated. Moreover, $M_{\text{Pl}}$ will denote the reduced Planck mass.

\section{Solar chameleons}
\label{sec:solarchameleons}

We begin by briefly describing the chameleon mechanism, before concisely reviewing the production of chameleons in the Sun (very closely following Paper~II).

\subsection{Chameleon screening}
\label{subsec:chameleon}

Several screening mechanisms, including the chameleon one, can be formulated in terms of scalar-tensor theories: a key feature of these theories, well known since the 1961 Brans-Dicke work~\cite{Brans:1961sx}, is the freedom to choose a frame, which determines how the scalar field couples to gravity and matter. Two of the most widely used frames are the Einstein frame (EF), where the gravitational action takes the canonical Einstein–Hilbert form but matter couples non-minimally to the scalar field, and the Jordan frame (JF), where matter fields are minimally coupled, at the price of a gravitational action which is not the standard one: the relation between the EF and JF metrics is controlled by the scalar degree of freedom, in our case the chameleon field.~\footnote{Whether EF and JF, and more generally all these frames, are physically equivalent, is still an open question, with potentially important implications for observables, especially in the semi-classical or quantum regimes (see e.g.\ Refs.~\cite{Magnano:1993bd,Capozziello:1996xg,Faraoni:1999hp,Faraoni:2006fx,Capozziello:2010sc,Steinwachs:2011zs,Ren:2014sya,Postma:2014vaa,Kamenshchik:2014waa,Domenech:2015qoa,Myrzakulov:2015qaa,Quiros:2015bfa,Sakstein:2015jca,vandeBruck:2015gjd,Banerjee:2016lco,Kamenshchik:2016gcy,Bahamonde:2016wmz,Pandey:2016unk,Jarv:2016sow,Mathew:2017lvh,Karam:2017zno,Azri:2018gsz,Rinaldi:2018qpu,Falls:2018olk,Nashed:2019yto,Giacomini:2020grc,Elizalde:2020icc,Oikonomou:2020oex,Bamonti:2021jmg,Copeland:2021qby,Racioppi:2021jai,Galaverni:2021jcy,Shtanov:2022wpr,Paliathanasis:2022tmt,Nojiri:2022ski,Paliathanasis:2022akr,Karciauskas:2022jzd,DeAngelis:2022qhm,Bamber:2022eoy,Odintsov:2022bpg,Schiavone:2022wvq,Oikonomou:2023dgu,Velasquez:2023jld,Diaz:2023tma,Luongo:2024opv,GiontiSJ:2023tgx,Seleim:2023enf,Belfiglio:2024swy,Galaverni:2025acu,Wang:2025ger} for works in this important direction). We note that, while Einstein and Jordan frames are the two most widely used frames, there is in fact an entire family of mathematically equivalent conformally related frames.} Screening mechanisms are particularly transparent in the EF: since matter fields are minimally coupled to the JF metric, in the EF the scalar field mediates an additional fifth force. Screening acts to dynamically suppress this fifth force in regions where gravity is well tested, such as high-density environments: essentially, this is achieved by dynamically decoupling the scalar field. In the chameleon case, in the EF this decoupling is achieved by giving the scalar a large effective mass, which makes the associated fifth force sufficiently short-ranged to escape detection (from the JF perspective, screening effectively ``restores'' General Relativity in these environments).

To be more concrete, let us consider the following action for gravity, which includes a real scalar field $\phi$ (hereafter the chameleon field), and SM fields (both radiation and matter), collectively denoted by $\psi$:
\begin{align}
S=&\int d^4x\sqrt{-g}\, \left [ \frac{M_{\text{Pl}}^2}{2}R-\frac{1}{2}\partial_{\mu}\phi\partial^{\mu}\phi-V_{\text{self}}(\phi) \right ] \nonumber\\
&+S_{\text{SM}} \left [ \widetilde{g}_{\mu\nu}; \psi \right ] \,.
\label{eq:actionchameleon}
\end{align}
where $g_{\mu\nu}$ and $\widetilde{g}_{\mu\nu}$ are the EF and JF metrics respectively, with $g \equiv \det(g_{\mu\nu})$, $M_{\text{Pl}}$ is the reduced Planck mass, $S_{\text{SM}}$ is the action for SM fields, and $V_{\text{self}}$ is a (bare) self-interacting potential for the chameleon field. A common choice for $V_{\text{self}}$ is related to the Ratra-Peebles inverse power-law potential, and is given by the following:
\begin{equation}
V_{\text{self}}(\phi) = \Lambda^4 \left ( 1 + \frac{\Lambda^n}{\phi^n} \right) \,,
\label{eq:v}
\end{equation}
where $\Lambda$ sets the energy scale of the potential, and $n$ is an integer, which in what follows we will assume to be positive. Values of $\Lambda$ for which the chameleon can play a role in the cosmic acceleration phenomenon are those close to the so-called ``dark energy scale'' $\Lambda_{\text{DE}} \simeq 2.4{\rm\,meV}$. A key role in screening mechanisms is played by the relation between EF and JF metrics. As shown by Bekenstein, because of General Relativity being invariant under general diffeomorphisms, the generic relation between these metrics takes the following form~\cite{Bekenstein:1992pj,Zumalacarregui:2010wj,Sebastiani:2016ras}:
\begin{equation}
\widetilde g_{\mu\nu} = \mathcal{A}^2(\phi, X)\,g_{\mu\nu} + \mathcal{B}^2(\phi, X)\,\partial_\mu\phi\partial_\nu\phi\,,
\label{eq:disformal}
\end{equation}
where $X \equiv -(1/2)(\partial_\mu\phi)(\partial^\mu\phi)$ is the chameleon kinetic term, and the functions $\mathcal{A}(\phi, X)$ and $\mathcal{B}(\phi, X)$ describe the conformal and disformal components of the transformation between the two metrics. In what follows, we make the following choice for these two functions:
\begin{align}
\mathcal{A}^2(\phi,X)&=1+2\beta_i\frac{\phi}{M_{\rm Pl}}\,,
\label{eq:a} \\
\mathcal{B}^2(\phi,X)&=\frac{2}{M_i^4}\,
\label{eq:b}
\end{align}
In Eqs.~(\ref{eq:a},\ref{eq:b}), $\beta_i$ and $M_i$ are species-dependent coefficients controlling respectively the conformal and disformal couplings of the chameleon to the $i$th species. Note that, while the $\beta_i$s are dimensionless, the $M_i$s have dimensions of energy. With the above choice we are dealing with the same ``generalized'' chameleon scalar-tensor theory considered in the earlier Papers~I and~II.~\footnote{Note that the ``vanilla'' chameleon model corresponds to the choice $\mathcal{A}(\phi,X) = \exp(\beta_m\phi/M_{\rm Pl})$, of which Eq.~(\ref{eq:a}) can be understood as a first-order Taylor expansion, which is appropriate since the field is always sub-Planckian. Moreover, note that in Eqs.~(\ref{eq:a},\ref{eq:b}) neither $\mathcal{A}$ nor $\mathcal{B}$ depend on $X$. This means that we are implicitly ignoring a possible kinetic-conformal coupling. See Ref.~\cite{Vagnozzi:2021quy} for further discussions on the reason for this choice.}

For our purposes, it is simplest to rewrite the action of Eq.~(\ref{eq:actionchameleon}) entirely in the EF, taking into account Eqs.~(\ref{eq:v}--\ref{eq:b}). This leads to the following effective action describing the coupling of the chameleon to matter fields and photons:
\begin{align}
S &= \int d^4x\sqrt{-g}\, \Biggl[ \frac{M_{\text{Pl}}^2}{2}R-\frac{1}{2}\partial_\mu\phi\partial^\mu\phi-V_{\text{eff}}(\phi) \Biggr. \nonumber \\
& \Biggl. +\frac{1}{M_{\gamma}^4}\partial_{\mu}\phi\partial_{\nu}\phi T_{\gamma}^{\mu\nu}+\frac{1}{M_i^4}\partial_{\mu}\phi\partial_{\nu}\phi T_{i}^{\mu\nu} \Biggr. \nonumber \\
& \Biggl. -\frac{1}{4}F^{\mu\nu}F_{\mu\nu}+A_{\mu}J^{\mu}+\mathcal{L}_m \Biggr] \,,
\label{eq:effective}
\end{align}
where $T_{i}^{\mu\nu}$ is the stress-energy tensor of matter species (excluding photons, discussed later), $F_{\mu\nu}=\partial_{\mu}A_{\nu}-\partial_{\nu}A_{\mu}$ is the photon field strength tensor, with $A_{\mu}$ and $J^{\mu}$ respectively the electromagnetic four-potential and four-current, whereas $\mathcal{L}_m$ is the non-electromagnetic part of the SM Lagrangian density, and $T_{\gamma}^{\mu\nu}$ is the electromagnetic stress-energy tensor, given by the following:
\begin{equation}
T^{\mu\nu}_\gamma = F^{\mu\alpha}{F^\nu}_\alpha-\frac{1}{4}g^{\mu\nu}F^{\alpha\beta}F_{\alpha\beta}\,.
\label{eq:tmunu}
\end{equation}
Finally, $V_{\text{eff}}$ is the \textit{effective} potential to which the chameleon responds in the presence of surrounding matter with total density $\rho_m$:
\begin{equation}
V_{\text{eff}}(\phi) = V_{\text{self}}(\phi)+\frac{\beta_m}{M_{\text{Pl}}}\rho_m\phi+\frac{\beta_\gamma}{4M_{\rm Pl}}\phi F^{\mu\nu}F_{\mu\nu}\,,
\label{eq:veff}
\end{equation}
where, considering matter composed of individual species with conformal couplings $\beta_i$ and densities $\rho_i$, the coupling $\beta_m$ is defined as $\beta_m \equiv \sum_i\beta_i\rho_i/\rho_m$. Crucially, although neither of the terms in Eq.~(\ref{eq:veff}) possesses a minimum, the resulting effective potential does possess a density-dependent minimum, which in turn determines the effective density-dependent mass of the chameleon. In the regime where its effective mass is sufficiently large compared to the relevant environmental scale, the chameleon adiabatically tracks the minimum of the potential. Under this assumption, which we will adopt in what follows, and ignoring the photon contribution from Eq.~(\ref{eq:veff}), the density-dependent chameleon field value and effective mass squared are given by the following~\cite{Khoury:2003aq}:~\footnote{The photon contribution can safely be ignored, as is usually done in the literature, unless one is in the presence of extremely strong magnetic fields, several orders of magnitude beyond the dynamical range which will be of interest to us in this work.}
\begin{align}
\phi(\rho) &= \left ( \frac{nM_{\text{Pl}}\Lambda^{4+n}}{\beta_m\rho} \right ) ^{\frac{1}{1+n}}\,,
\label{eq:phimin} \\
m_{\text{eff}}^2(\rho) &\approx n(1+n)\Lambda^{4+n} \left ( \frac{\beta_m\rho}{nM_{\text{Pl}}\Lambda^{4+n}} \right ) ^{\frac{2+n}{1+n}}\,.
\label{eq:meff}
\end{align}
The key point is that Eq.~(\ref{eq:meff}) increases with increasing $\rho_m$. This implies that the effective chameleon mass increases in dense environments, such as in the Solar System. It follows that the range of the associated fifth force decreases, a feature which allows the model to pass precision local tests of gravity. On the other hand, the field can remain ultra-light on cosmological scales, leaving open the possibility that it plays a role in the cosmic acceleration phenomenon.

In what follows, we neglect the disformal coupling to photons, i.e.\ the $\partial_{\mu}\phi\partial_{\nu}\phi T_{\gamma}^{\mu\nu}/M_{\gamma}^4$ term in Eq.~(\ref{eq:effective}), therefore formally taking the $M_{\gamma} \to \infty$ limit. The first reason is that current limits on $M_{\gamma}$ are already rather stringent~\cite{Brax:2013nsa}. More importantly, as observed earlier in Paper~I, we require $M_{\gamma} \gg {\cal O}({\text{keV}})$ in order to remain consistent with stellar cooling constraints from energy loss in horizontal branch stars. Under this assumption, solar production is then dominated by the conformal channel, i.e.\ the term $\mathcal{L}_{\phi\gamma\gamma} \propto \beta_\gamma\phi F^2$ in Eq.~(\ref{eq:effective}), which can lead to conversion of photons into chameleons in the presence of magnetic fields and viceversa, as we will discuss later. Summing up, the effective action we will consider in the present work is given by the following (for clarity we work in the Einstein frame):
\begin{align}
S &= \int d^4x\sqrt{-g}\, \Biggl[ \frac{M_{\text{Pl}}^2}{2}R-\frac{1}{2}\partial_\mu\phi\partial^\mu\phi-\Lambda^4 \left ( 1 + \frac{\Lambda^n}{\phi^n} \right) \Biggr. \nonumber \\
& \Biggl. +\frac{\beta_m}{M_{\text{Pl}}}\rho_m\phi+\frac{\beta_\gamma}{4M_{\rm Pl}}\phi F^{\mu\nu}F_{\mu\nu}+\frac{1}{M_i^4}\partial_{\mu}\phi\partial_{\nu}\phi T_{i}^{\mu\nu} \Biggr. \nonumber \\
& \Biggl. -\frac{1}{4}F^{\mu\nu}F_{\mu\nu}+A_{\mu}J^{\mu}+\mathcal{L}_m \Biggr] \,,
\label{eq:effectivefinal}
\end{align}
where we remark once again that $g_{\mu\nu}$ is the EF metric tensor. In Eq.~(\ref{eq:effectivefinal}), the terms in the first row define the gravitational sector and the chameleon self-coupling (which ultimately controls its dynamics on cosmological scales). The terms in the second row instead characterize the chameleon conformal couplings to matter and photons, with coupling strengths $\beta_m$ and $\beta_{\gamma}$ respectively, and disformal couplings to matter, controlled by the scales $M_i$. Finally, the terms in the third row essentially describe the SM Lagrangian. For what concerns the interactions of chameleons with matter fields, in what follows we shall mostly be interested in interactions with electrons, since we will use electron recoil data from XENONnT. Therefore, the matter couplings of interest to us will be the conformal and disformal couplings to electrons, whose associated coupling strength and energy scale we will denote by $\beta_e$ and $M_e$ respectively. In short, for the purposes of our work the model is therefore specified by five parameters: $\Lambda$, $n$, $\beta_e$, $\beta_{\gamma}$, and $M_e$ (although, as we will see later, only $\beta_{\gamma}$ and $M_e$ play a major role).

\subsection{Solar production of chameleons}
\label{subsec:solarproduction}

Because of its hot and dense environment, the presence of a strong magnetic field, and its proximity to Earth, the Sun provides a natural setting for the production of light particles~\cite{Raffelt:1996wa} (see e.g.\ Refs.~\cite{Redondo:2013lna,Giannotti:2015kwo,Giannotti:2017hny,Chang:2018rso,Budnik:2019olh,Carenza:2020zil,OHare:2020wum,Lucente:2020whw,Pallathadka:2020vwu,Dev:2020jkh,Carenza:2020cis,Edwards:2020afl,Caputo:2021kcv,Caputo:2021eaa,Calore:2021klc,Fischer:2021jfm,Caputo:2021rux,Berlin:2021kcm,Caputo:2022mah,Lucente:2022vuo,Balaji:2022noj,Akita:2022etk,Fiorillo:2022cdq,Bottaro:2023gep,Lella:2023bfb,Hoof:2023jol,Carenza:2023lci,Akita:2023iwq,Caputo:2025aac,Carenza:2025uib,Fiorillo:2025yzf,Fiorillo:2025sln,Fiorillo:2025zzx,Fiorillo:2025gnd} for examples of works on stellar production of new ultralight particles). This includes the potential production of chameleon scalars. We now briefly review our calculation of the spectrum of solar chameleons, closely following Paper~II. The underlying assumption is the existence of a scalar field conformally coupled to photons and with a mass term $\mathcal{L} \propto -m^2\phi^2$, suitable for either a fixed-mass scalar or a chameleon field, with $m$ replaced by the density-dependent mass in Eq.~(\ref{eq:meff}). We include two channels: Primakoff production in the electric fields of electrons and ions, and magnetic conversion in the bulk magnetic field of the Sun. We now briefly review each of these channels in turn.

\subsubsection{Primakoff production}
\label{subsubsec:primakoff}

We begin by considering Primakoff production within the electric fields of electrons and ions in the solar plasma. Not only is this process the dominant one for production of solar chameleons but, unlike magnetic conversion, it is only marginally affected by astrophysical uncertainties, as it depends primarily on the very well-measured solar temperature, density, and chemical composition. The relevant Lagrangian terms are the following:
\begin{equation}
\mathcal{L} \supset -\frac{\beta_{\gamma}}{4 M_{\rm Pl}} \phi F_{\mu\nu}F^{\mu\nu} - e A_{\mu}\bar{\psi}\gamma^{\mu}\psi\,,
\label{eq:primakoff}
\end{equation}
which allow for conversion of photons into chameleons through their interactions with electrons and ions: $\gamma + Ze \to Ze + \phi$. The resulting spectrum, or more precisely number of chameleons produced per unit time per unit energy, which we refer to d$\dot{N}_P/{\rm d}\omega$, was computed by some of us in Paper~II and takes the following form:
\begin{equation}
\frac{d\dot{N}_P}{d\omega}=\frac{\beta_{\gamma}^2\alpha}{8\pi M_{\text{Pl}}^2}\int_0^{R_{\odot}}\frac{{\rm d}r\,r^2}{e^{\omega/T}-1}\frac{\omega^2k_{\phi}}{k_\gamma}\mathcal{I}(u,v)\sum_iZ_i^2n_i\,,
\label{eq:productionprimakoff}
\end{equation}
where $\alpha$ is the fine-structure constant, $\omega$ is the energy of the produced chameleon (which is equal to the energy of the pre-conversion photon, since energy is conserved in the photon-chameleon conversion process, while the momentum mismatch is absorbed by the solar plasma), $k_{\phi}$ and $k_{\gamma}$ are respectively the (density-dependent) chameleon 3-momentum and photon 3-momentum. These are given by the dispersion relations $k_{\phi}=\sqrt{\omega^2-m_{\phi}^2(\rho)}$ and $k_{\gamma}=\sqrt{\omega^2-m_{\gamma}^2}$, where $m_{\gamma}$ is the plasma-induced effective photon mass, given by the plasma frequency $\omega_{\text{pl}}=\sqrt{4\pi\alpha\, n_e/m_e}$, where $n_e$ and $m_e$ are the electron number density and mass respectively. In Eq.~(\ref{eq:productionprimakoff}), the sum runs over all charged species in the plasma, and the integral is performed over the radial coordinate of the Sun, with $R_{\odot}$ being the radius of the Sun, while the function $\mathcal{I}(u,v)$ is defined as follows:
\begin{equation}
\begin{split}
&\mathcal{I}(u,v)\equiv\int_{-1}^{+1}{\rm d}x\,\frac{1-x^2}{(u-x)(v+u-x)}\\
&= \frac{(u+v)^2-1}{v} \ln \left ( \frac{u+v+1}{u+v-1} \right) - \frac{u^2-1}{v} \ln \left( \frac{u+1}{u-1} \right) - 2\,,
\end{split}
\label{eq:iuv}
\end{equation}
with the auxiliary variables $u$ and $v$ being given by the following expressions:
\begin{equation}
u=\frac{k_{\gamma}}{2k_{\phi}}+\frac{k_{\phi}}{2k_{\gamma}}\,, \quad v=\frac{\kappa^2}{2k_{\gamma}k_{\phi}}\,,
\end{equation}
where the Debye screening scale $\kappa$ controls modifications to the photon propagator in the non-degenerate plasma due to Debye screening, and is defined as follows~\cite{Raffelt:1996wa}:
\begin{equation}
\kappa = \frac{4\pi A}{T}\sum_iZ_i^2n_i\,,
\label{eq:debye}
\end{equation}
with the sum extending over all the charged particles present in the Sun, whose atomic numbers and number densities are given by $Z_i$ and $n_i$ respectively.

It is important to note that the spectrum given in Eq.~(\ref{eq:productionprimakoff}) scales as $\beta_{\gamma}^2$, the reason being that Primakoff production is mediated by the $\phi F^{\mu\nu}F_{\mu\nu}$ interaction, whose coupling strength is $\beta_{\gamma}/M_{\text{Pl}}$. Given that the production rate is proportional to the square of the associated amplitude, the former scales as $\beta_{\gamma}^2$, a consideration which will turn out to be particularly important when we discuss how the resulting electron recoil event rate in the XENONnT detector depends on the five chameleon model parameters.

\subsubsection{Magnetic conversion}
\label{subsubsec:magnetic}

The second channel we include is magnetic conversion in the bulk solar magnetic field (while Paper~I originally considered only conversion in the tachocline, Paper~II included the full magnetic field profile). As in the axion case~\cite{Caputo:2020quz,Guarini:2020hps}, the computation is based on the thermal field theory framework (an alternative kinetic approach results in the same production rate, as discussed in Paper~II). It can be shown that the Lagrangian in Eq.~(\ref{eq:effectivefinal}) contains the following term of interest for this process:
\begin{equation}
\mathcal{L} \supset \frac{\beta_\gamma}{M_{\rm Pl}} \mathbf{B} \cdot (\nabla \phi \times \mathbf{A})\,,
\label{eq:magnetic}
\end{equation}
which allows for conversion of photons into chameleons in the presence of an external magnetic field. The resulting spectrum, d$\dot{N}_B/{\rm d}\omega$, was computed by some of us in Paper~II, and is given by the following expression:
\begin{align}
&\frac{{\rm d}\dot{N}_B}{{\rm d}\omega} = \nonumber \\
&\frac{2\beta_{\gamma}^2}{\pi M_{\text{Pl}}^2}\int_0^{R_{\odot}}{\rm d}r\,r^2B_{\perp}^2(r)\frac{\omega(\omega^2-m^2)^{3/2}}{(m_{\gamma}^2 - m^2)^2+(\omega\Gamma_{\gamma})^2}\frac{\Gamma_{\gamma}}{e^{\omega/T}-1}\,,
\label{eq:productionmagnetic}
\end{align}
where, without loss of generality, we have aligned the magnetic field along the $\hat{z}$-direction, so $B_{\perp}^2=B_x^2+B_y^2$ denotes the magnetic field component perpendicular to the propagation direction. Here, $m_{\gamma}$ is the plasma-induced effective photon mass discussed in Sec.~\ref{subsubsec:primakoff}, whereas $\Gamma_{\gamma}(\omega,r)$ accounts for a number of photon production and absorption processes within the solar medium. Following Paper~II, given the relatively high energies we are interested in, we assume that only the Thomson and free-free contributions are relevant, so $\Gamma_{\gamma}$ takes the following form:
\begin{align}
\Gamma_{\gamma}=\frac{64\pi^2\alpha^3}{3m_e^2\omega^3}\sqrt{\frac{m_e}{2\pi T}}n_e \! \left (\!1\!-\!e^{-\omega/T}\!\right ) \! \sum_iZ_i^2n_iF_i +\frac{8\pi\alpha^2}{3m_e^2}n_e\,,
\label{eq:gammagamma}
\end{align}
where the sum in principle runs over all ion species in the Sun, and the quantity $F_i$ denotes the thermally-averaged Gaunt factor for species $i$, which we take from Ref.~\cite{Chluba:2019ser}. Since the number abundances $n_i$ of heavier elements in the Sun, such as oxygen and iron, are strongly suppressed relative to hydrogen and helium, their contributions provide only a few-percent correction to the hydrogen term in this weighted sum. The dominant sub-leading contribution is thus that of helium. The Gaunt factors $F_i$ are of order unity and do not qualitatively alter this conclusion. As in Paper~II, we thus only include hydrogen and helium in the sum.

From Eq.~(\ref{eq:productionmagnetic}) we see that a key role in the magnetic conversion spectrum is played, unsurprisingly, by the radial profile of the solar magnetic field, $B(r)$. For this reason, unlike its Primakoff counterpart discussed earlier, the magnetic conversion channel suffers from larger astrophysical uncertainties, due to the less well known solar magnetic field (in particular within the radiative zone). In what follows, we adopt the same model for the magnetic field used in Paper~II, whose properties we briefly summarize in Appendix~\ref{app:solarmodel}, to which we refer the reader for further details.

\subsubsection{Complete spectrum}
\label{subsubsec:completespectrum}

\begin{figure}[htbp]
\centering
\includegraphics[width=0.9\linewidth]{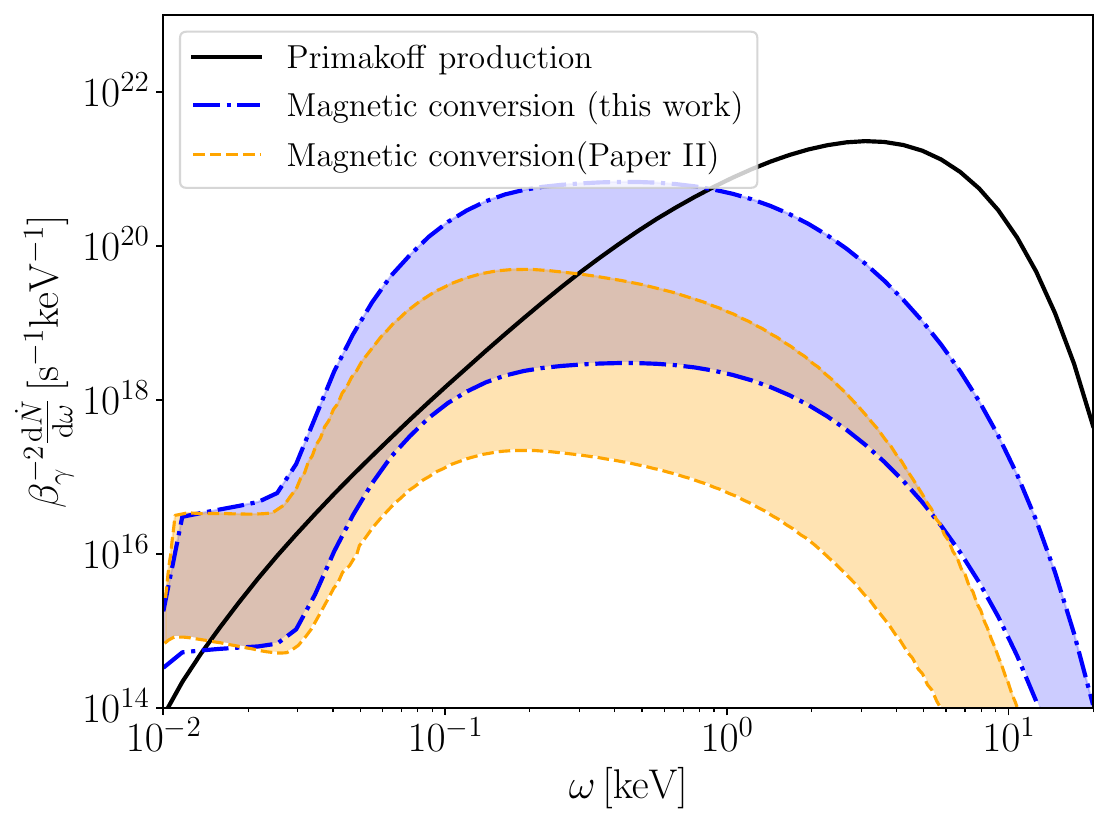}
\caption{Comparison of the two components of the solar chameleon spectrum: Primakoff production from transverse photons in the electric fields of electrons and ions (black solid curve), and magnetic conversion from the solar bulk magnetic field (blue dash-dotted curve and blue band). The width of the blue band reflects the uncertainty on the strength of the solar magnetic field (discussed in Appendix~\ref{app:solarmodel}). The yellow dashed curve and associated band correspond instead to the magnetic conversion component calculated in Paper~II, updated here (to the blue dash-dotted curve and associated band) to account for a few issues in the earlier code, as discussed in Sec.~\ref{subsubsec:completespectrum}. The spectrum is computed assuming the AGSS09 solar model, while fixing the chameleon parameters to $\beta_m=10^2$, $\Lambda=2.4\,{\text{meV}}$, and $n=1$. Note that the spectrum has been normalized by $\beta_{\gamma}^2$ to factor out the dependence on the chameleon-photon coupling strength, see Eqs.~(\ref{eq:productionprimakoff},\ref{eq:productionmagnetic}).}
\label{fig:spectrum}
\end{figure}

We show the complete spectrum of solar chameleons in Fig.~\ref{fig:spectrum}, where we have separated the contributions from Primakoff production in the electric fields of electrons and ions (black curve), and magnetic conversion (blue band). We see that the former dominates at larger energies, whereas the latter is the dominant contribution at lower energies. We can anticipate that, at the ${\cal O}({\text{keV}})$ recoil energies of interest to our analysis, the dominant contribution will be the Primakoff one: this expectation will be explicitly confirmed later. The width of the blue band reflects uncertainties in the solar magnetic field strength, discussed in more detail in Appendix~\ref{app:solarmodel}.

The magnetic conversion contribution to the spectrum computed here is slightly different (and in particular larger) compared to the one computed in Paper~II, owing to a few coding bugs we identified while carrying out the current work. The result reported in Paper~II is given by the yellow band in  Fig.~\ref{fig:spectrum}: we see that the updated magnetic conversion contribution to the spectrum is up to a couple of orders of magnitude stronger than suggested by the previous calculation, while still remaining subdominant relative to the Primakoff contribution at sufficiently high energies. The main issue we identified in the previous code concerns the implementation of Eq.~(\ref{eq:productionmagnetic}): in the \texttt{C++} code of Paper~II, the $(\omega^2-m^2)^{3/2}$ contribution at the numerator had been incorrectly coded up as ${\text{pow}}(\omega^2-m^2,3/2)$ instead of ${\text{pow}}(\omega^2-m^2,3.0/2.0)$, effectively resulting in the exponent being treated as the integer 1, rather than the float 1.5. This was the main source of error in the previous calculation, and led to the magnetic conversion component being underestimated by up to a couple of orders of magnitude. Other minor errors, which nevertheless are completely subdominant compared to this one, concern an incorrect reading of the Gaunt factor arrays in Eq.~(\ref{eq:gammagamma}), as well as a missing factor of $2/3$ in the calculation of the perpendicular component of the magnetic field $B_{\perp}$ in Eq.~(\ref{eq:productionmagnetic}).

All these issues have been corrected in the present work. We note, however, that their impact on XENONnT constraints on the chameleon parameters is rather minimal, given that the Primakoff contribution is completely dominant at the energies of interest.

We close with an important caveat. As discussed in Paper~II, the production of solar chameleons receives an additional contribution from longitudinal plasmons, which can contribute to Primakoff production. A dedicated calculation of this contribution to the solar chameleon spectrum has not yet been performed, and we therefore do not include it in the present work. Therefore, the spectrum shown in Fig.~\ref{fig:spectrum} should be regarded as a lower limit to the complete solar chameleon spectrum, since the plasmon contribution would only enhance it. Consequently, all limits we derive on the chameleon parameters are conservative, in the sense that including the plasmon contribution would only lead to tighter limits.


\section{Methodology}
\label{sec:methodology}

We now begin by briefly discussing theoretical expectations for the signal in the XENONnT detector resulting from interactions between solar chameleons and electrons, before outlining the statistical methods utilized to compare these theoretical expectations against the measured event rate.

\subsection{Event rate in the XENONnT detector}
\label{sec:event}

The XENON experiment, based at the Laboratori Nazionali del Gran Sasso under the Gran Sasso d'Italia massif in Italy, is primarily designed as a DM direct detection experiment. In particular, its main goal is to search for rare nuclear or electron recoils arising from interactions of hypothetical DM particles, such as weakly interacting massive particles or axion-like particles, using a dual-phase liquid-gas xenon time projection chamber. XENONnT, which began taking science data in September 2021, represents the latest and most sensitive stage of the program~\cite{XENON:2024wpa}. Compared to its predecessor XENON1T, it features a significantly larger active liquid xenon mass of about $5.9$ tonnes, while achieving substantially lower background levels thanks to an extensive material screening (radioassay) campaign and a new high-flow radon removal system~\cite{XENON:2021mrg}. In 2020, XENON1T reported an unexpected excess of electronic recoil events at recoil energies $E_R \lesssim 7\,{\rm keV}$~\cite{XENON:2020rca}. While its origin was never conclusively established, despite the possibility for trace amounts of tritium contamination, several beyond the SM possibilities for the excess were studied, ranging from solar axions to a number of dark matter candidates (see e.g.\ Refs.~\cite{Takahashi:2020bpq,Kannike:2020agf,Alonso-Alvarez:2020cdv,Boehm:2020ltd,Fornal:2020npv,Su:2020zny,Chen:2020gcl,Choi:2020udy,DiLuzio:2020jjp,An:2020bxd,Bloch:2020uzh,Zu:2020idx,Lindner:2020kko,Gao:2020wer,Shoemaker:2020kji,Farzan:2020llg,Zu:2020bsx,Adams:2020ejt,Aboubrahim:2020iwb,Buttazzo:2020vfs} for an incomplete list of examples). Another explanation, studied by some of us in Paper~I, is that the XENON1T excess may have been due to solar chameleons, opening a concrete pathway towards direct detection of DE in terrestrial DM detectors~\cite{Vagnozzi:2021quy}. No analogous excess is present in the current XENONnT electron recoil dataset~\cite{XENON:2022ltv}, allowing instead for a robust exclusion analysis in the chameleon parameter space. This is particularly valuable, as it provides limits that are completely independent of all other existing constraints on chameleons.

As shown in Fig.~\ref{fig:spectrum}, solar chameleons are produced with typical energies in the ${\cal O}({\text{keV}})$ range. Assuming they manage to escape the solar interior (see discussion below), these particles propagate nearly freely to the Earth, where they can interact with electrons in the XENONnT detector, primarily via absorption. The ``raw'' differential event rate per unit recoil energy, before accounting for energy resolution and detector efficiency effects, is given by the following:
\begin{align}
\frac{{\rm d}R_0(E_R)}{{\rm d}E_R}&=\sigma_{\phi e}\frac{{\rm d}\Phi_{\rm Earth}}{{\rm d}E_R} \nonumber \\
&=\sigma_{\phi e}\frac{1}{4\pi D^2} \left ( \frac{{\rm d}\dot{N}_{P}}{{\rm d}E_R}+\frac{{\rm d}\dot{N}_{B}}{{\rm d}E_R} \right ) \,,
\label{eq:dr0der}
\end{align}
where $\sigma_{\phi e}$ denotes the chameleon-electron interaction cross-section per unit target mass (hereafter referred to as cross-section for simplicity, and discussed in more detail below) and ${\rm d}\Phi_{\rm Earth}/{\rm d}E_R$ is the solar chameleon flux (number of chameleons produced per unit time per unit energy) incident at Earth. The latter is obtained by summing the Primakoff and magnetic conversion production rates in the Sun, namely ${\rm d}\dot{N}_{P}/{\rm d}E_R$ and ${\rm d}\dot{N}_{B}/{\rm d}E_R$ in Eqs.~\eqref{eq:productionprimakoff} and~\eqref{eq:productionmagnetic} respectively, and diving by $4\pi D^2$, with $D$ the distance between the Earth and the Sun. We express the resulting raw differential event rate in units of ${\text{ton}}^{-1}\,{\text{yr}}^{-1}\,{\text{keV}}^{-1}$. To obtain the full theoretical expectation for the differential event rate, we need to fold in detector response effects. The resulting theoretical (hence the $_{\text{th}}$ subscript) differential event rate is given by the following expression:
\begin{equation}
\left ( \frac{{\rm d}R}{{\rm d}E_R} \right )_{\rm th}=\epsilon(E_R)\int {\rm d}E\,\frac{{\rm d}R_0(E)}{{\rm d}E}\Theta(E_R-E)\,.
\label{eq:drder}
\end{equation}
where $\epsilon(E)$ is the XENONnT detection efficiency function (see Fig.~1 of Ref.~\cite{XENON:2022ltv}), and $\Theta(E)$ characterizes the detector energy resolution.~\footnote{In XENON1T the resolution was modeled via a Gaussian distribution with energy-dependent width, parametrized as $\sigma(E)=a\sqrt{E}+bE$, with $a=(0.310\pm 0.004)\,\sqrt{\text{keV}}$ and $b=0.0037 \pm 0.0003$~\cite{XENON:2020rca}. In XENONnT, the resolution has instead been shown to be better captured by a skew Gaussian model (see e.g.\ Fig.~23 of Ref.~\cite{XENONCollaboration:2024bil}). Nevertheless, we have verified that the differences between adopting a Gaussian versus skew Gaussian are minimal, and we therefore stick to the Gaussian approximation in this work. The detection efficiency function has been taken from the publicly available repository \href{https://zenodo.org/records/7311940}{zenodo.org/records/7311940}.} The theoretical differential event rate in Eq.~(\ref{eq:drder}) is the quantity which can be directly compared to the experimentally measured event rate to constrain the chameleon parameters.

As alluded to earlier, detection of solar chameleons in experiments such as XENONnT proceeds through absorption by electrons. In fact, this is conceptually similar (although with important differences) to the way axions and axion-like particles can be detected in the same experiments, through a process known as the axio-electric effect. In our case, the two relevant Lagrangian terms for detection are the conformal and disformal couplings of chameleons to electrons, i.e.\ the first and third term in the second row of Eq.~(\ref{eq:effectivefinal}), controlled by the couplings $\beta_e$ and $M_e$ respectively (recall that we are now specializing to electrons as the matter fields of interest, hence the subscript $_e$). Both terms contribute to the cross-section $\sigma_{\phi e}$, which was extensively discussed in Paper~I (see especially Appendix~C thereof), and is given by the following:
\begin{align}
\sigma_{\phi e}(E_R)&=\sigma_{\phi e,{\text{dis}}}(E_R)+\sigma_{\phi e,{\text{conf}}}(E_R) \nonumber \\
&=N_{\rm Xe}\,\frac{m_e^2E_R^4}{8\pi^2M_e^8}+\frac{\beta_e^2E_R^2}{2\pi\alpha M_{\text{Pl}}^2}\sigma_{\text{photo}}\,,
\label{eq:sigmaphie}
\end{align}
where $N_{\text{Xe}}=4.6\times 10^{27}\,{\text{ton}}^{-1}$ is the number of xenon atoms per ton of material. The two terms $\sigma_{\phi e,{\text{dis}}}$ and $\sigma_{\phi e,{\text{conf}}}$ refer to the contributions from the disformal and conformal couplings, respectively, and do not interfere in the calculation of the overall amplitude, therefore giving rise to two separate, clearly distinguished terms. Finally, in Eq.~(\ref{eq:sigmaphie}), $\sigma_{\text{photo}}$ and $m_e$ denote the (energy-dependent) photo-electric cross-section from Ref.~\cite{Veigele:1973tza} and electron mass respectively.~\footnote{In Paper~I, the above cross-section was referred to as ``\textit{chameleo-electric}'' cross-section, as it is the chameleon analogue of the photo-electric and axio-electric effects, where electrons absorb photons and axions respectively, recoiling with higher energies. Although this is simply a matter of terminology and not of substance, the analogy should technically speaking only be extended to the conformal term, and therefore the proper chameleo-electric cross-section should be $\sigma_{\phi e,{\text{conf}}}$. In what follows, to avoid this issue, we will refrain from referring to this cross-section and the associated process through a specific name.} The above cross-section has been derived under the assumption that the chameleon mass in the detector is much lighter than the two other relevant energy scales in the problem: $E_R$ and $m_e$. For all intents and purposes, the chameleon can therefore be considered (quasi)-massless (see Paper~I for further discussions). The reason for this being a good approximation is that, in chambers of typical size $R$, the chameleon settles into a standing wave configuration set by a resonance condition involving the size of the cavity, such that the associated effective mass is of order $R^{-1}$~\cite{Khoury:2003aq,Khoury:2003rn,Brax:2007hi,Brax:2012gr}. In the case of XENONnT, where the size of the time projection chamber is of order ${\cal O}({\text{m}})$, the associated chameleon effective mass is of order ${\cal O}(10^{-7}\,{\text{keV}})$, thus well below the other two relevant energy scales, justifying the approximation of extremely light chameleon.

It is interesting to note that the disformal part of the cross-section features an energy scaling of $E_R^4$~\cite{Vagnozzi:2021quy}. This is qualitatively very different from the axion case, where the axio-electric cross-section is an appropriately rescaled version of the conformal part of the cross-section, with overall coupling $g_{ae}^2$ (where $g_{ae}$ is the axion-electron coupling), i.e.\ the axion counterpart of $\beta_e^2$ appearing in our case. The reason for such a different scaling lies in the higher-derivative nature of the disformal operator. This, combined with the fact that the disformal channel will end up dominating the detection cross-section as we will see later, leads to a qualitatively rather different phenomenology for chameleons compared to axions.

The detection cross-sections for chameleons and axion come with very different dependence on the target materials. For solar axions the event rate is governed by the axio-electric effect, which approximately tracks the photoelectric cross-section and therefore grows quickly with the target atomic number. In contrast, for solar chameleons, the dominant disformal contribution is nearly target-independent, aside from experiment-specific effects such as energy thresholds and efficiencies. Therefore, for targets with very different atomic numbers, the axion hypothesis predicts rate differences of orders of magnitude, while the chameleon hypothesis predicts comparable rates once normalized appropriately. As a consequence, one could plausibly discriminate between the two scenarios without requiring percent-level precision in the relative normalization, but simply sufficient statistics (and control of systematics) to distinguish an $\mathcal{O}(1)$ relative rate from an $\mathcal{O}\left(10^2-10^3\right)$ one.

A comment is in order before moving on. As anticipated earlier, all the above considerations hinge upon the chameleons being able to escape from the Sun, and make it through all relevant materials they encounter between the Sun and the detector. Chameleons can traverse a barrier with local matter density $\rho$ and temperature $T$ provided that $m_{\text{eff}}(\rho) \ll T$. As stated earlier, we focus on $n>0$ chameleons, whose effective mass is a monotonically increasing function of local density. Therefore, for our purposes it is sufficient to ensure that chameleons make it through the densest material they encounter on their path. In our case this is the solar core, where $\rho_{\text{core}} \approx 150\,{\text{g}}/{\text{cm}}^3 \approx 6.5\times 10^{20}\,{\text{eV}}^4$ and $T_{\text{core}} \approx 1.5\times 10^7\,{\text{K}} \approx 1.3\times 10^3\,{\text{eV}}$. In fact, all the other materials subsequently traversed by solar chameleons, such as limestone or lead, are all significantly less dense than the core of the Sun.

As we clearly see from Eq.~(\ref{eq:meff}), the effective mass of the chameleon is controlled, aside from the local density $\rho$, by the three model parameters $\Lambda$, $\beta_m$, and $n$. The requirement that the chameleon is able to escape the Sun therefore sets constraints on the regions of $\Lambda$-$\beta_m$-$n$ parameter space which can serve as viable detection benchmarks. For concreteness, we start from the most widely studied (and constrained) case of $n=1$ chameleons. In this case, Eq.~(\ref{eq:meff}) reduces to the following expression:
\begin{equation}
m_{\text{eff}}(\rho) \approx \sqrt{2}\Lambda^{-\frac{5}{4}} \left ( \frac{\beta_m\rho}{M_{\rm Pl}} \right ) ^{\frac{3}{4}}\,.
\label{eq:meffn1}
\end{equation}
Focusing on the solar core, the ``escape condition'' can then be expressed as follows:
\begin{equation}
 \sqrt{2}\Lambda^{-\frac{5}{4}} \left ( \frac{\beta_m\rho_{\text{core}}}{M_{\text{Pl}}} \right ) ^{\frac{3}{4}} \ll T_{\text{core}} \implies \Lambda \gg \left ( \frac{4\beta_m^3\rho_{\text{core}}^3}{T_{\text{core}}^4M_{\text{Pl}}^3} \right ) ^{\frac{1}{5}}\,.
\label{eq:escapecore}
\end{equation}
We see that, for a given $\widetilde{\beta}_m$, Eq.~(\ref{eq:escapecore}) defines a ``critical value'' $\Lambda_{\text{crit}}^{(n=1)}$ given by the following:
\begin{equation}
\Lambda_{\text{crit}}^{(n=1)} \equiv \left ( \frac{4\widetilde{\beta}_m^3\rho_{\text{core}}^3}{T_{\text{core}}^4M_{\text{Pl}}^3} \right ) ^{\frac{1}{5}} \simeq 4.8 \times 10^{-7}\widetilde{\beta}_m^{\frac{3}{5}}\,{\text{eV}} \,,
\label{eq:lcritn1}
\end{equation}
such that for $\Lambda \gtrsim \Lambda_{\text{crit}}^{(n=1)}$ the chameleon can escape the solar core, or conversely is trapped for $\Lambda \lesssim \Lambda_{\text{crit}}^{(n=1)}$. Strictly speaking, the escape condition $m_{\text{core}} \lesssim T_{\text{core}}$ should not be a sharp transition but a gradual suppression effect. Nevertheless, treating it as a threshold criterion is sufficient for our purpose of identifying representative regions in $(\beta_m,\Lambda)$ parameter space which can (or cannot) serve as interesting benchmarks.

In Fig.~\ref{fig:chameleonescape}, focusing on $n=1$ chameleons, we show the regions of $\beta_m$-$\Lambda$ parameter space where chameleons can either escape the solar core (unshaded region above the solid black line), or remain trapped (light gray shaded region below the solid black line). The boundary between the two regions is set by the condition provided in Eq.~(\ref{eq:lcritn1}). On the same plane, we show the regions excluded by atom interferometry, levitated force sensors, and torsion balance experiments~\cite{Wagner:2012ui, Jenke:2014yel, Burrage:2017qrf, Yin:2022geb, MICROSCOPE:2022doy}. We note that levitated force sensors reportedly close the remaining window in $\beta_m$ parameter space, but only for $n=1$ chameleons~\cite{Yin:2022geb}. On the same plot, we also consider the theoretically interesting benchmark $n=4$, which we take as a representative steeper inverse-power case which has been studied in the literature (including in Paper~II). In this case, a calculation completely analogous to that carried out earlier for $n=1$ chameleons leads to the following expression for the ``critical value'' $\Lambda_{\text{crit}}^{(n=4)}$:
\begin{equation}
\Lambda_{\text{crit}}^{(n=4)} \equiv \left ( \frac{\sqrt{3125}\widetilde{\beta}_m^3\rho_{\text{core}}^3}{2T_{\text{core}}^5M_{\text{Pl}}^3} \right ) ^{\frac{1}{4}} \simeq 3.5 \times 10^{-9}\widetilde{\beta}_m^{\frac{3}{4}}\,{\text{eV}} \,,
\label{eq:lcritn4}
\end{equation}
In this case, the region of parameter space where $n=4$ chameleons can escape the solar core is given by the region above the dash-dotted black line, whereas the dark gray shaded region below the dash-dotted black line is that where chameleons remain trapped. The boundary between the two regions is set by the condition provided in Eq.~(\ref{eq:lcritn4}). We note that the levitated force sensor constraints do not apply to the $n=4$ case, whereas atom interferometry and torsion balance limits (not shown here to avoid overcrowding the figure) weaken relative to the $n=1$ case.

\begin{figure}[!ht]
\centering
\includegraphics[width=0.9\linewidth]{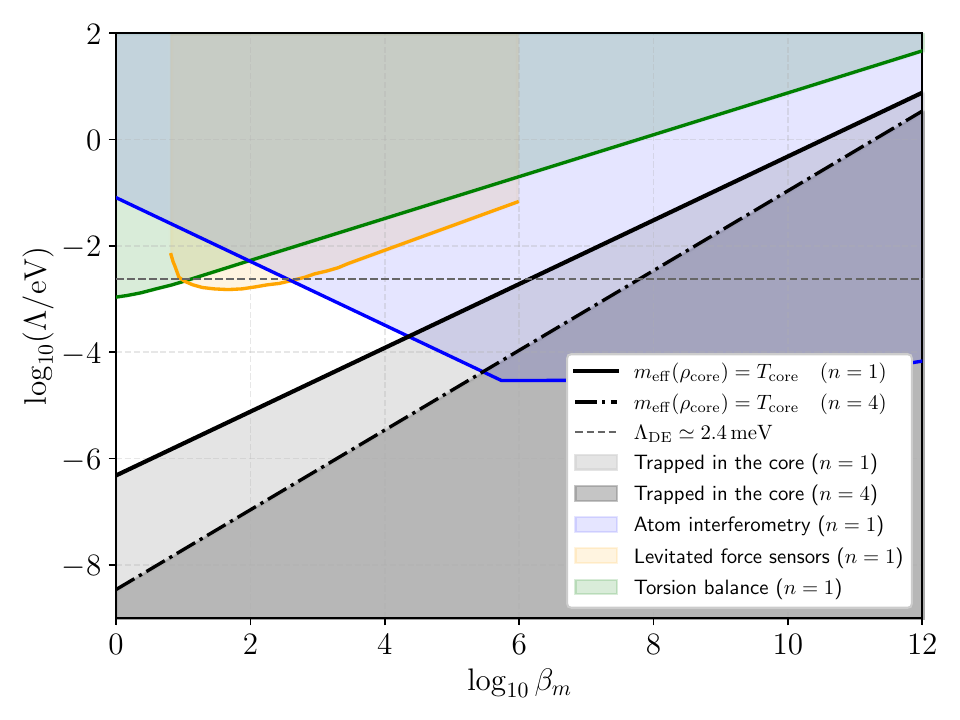}
\caption{Regions of $\beta_m$-$\Lambda$ parameter space where solar chameleons can escape the solar core, or conversely remain trapped. For $n=1$ chameleons the two regions are those above the solid black line and the light gray shaded region below the solid black line respectively, whereas for $n=4$ chameleons these are instead the region above the dash-dotted black line and the dark gray shaded region below the dash-dotted black line respectively. The solid black and dash-dotted black lines correspond to the boundaries defined by the ``critical values'' $\Lambda_{\text{crit}}^{(n=1)}(\beta_m)$ and $\Lambda_{\text{crit}}^{(n=4)}(\beta_m)$ given by Eqs.~(\ref{eq:lcritn1},\ref{eq:lcritn4}), where the effective mass of the ($n=1$ or $n=4$) chameleon in the solar core is equal to the temperature of the latter, $m_{\text{eff}}(\rho_{\text{core}})=T_{\text{core}}$. The blue, orange, and green shaded regions are those excluded by atom interferometry, levitated force sensors, and torsion balance experiments respectively, for $n=1$ chameleons. For $n=4$ chameleons the atom interferometry and torsion balance limits (not shown here to avoid overcrowding the figure) are weaker, and the levitated force sensor constraints do not apply. The horizontal dashed line corresponds to the DE scale $\Lambda \simeq 2.4\,{\text{meV}}$.}
\label{fig:chameleonescape}
\end{figure}

\subsection{Analysis}
\label{subsec:analysis}

To mirror the analysis carried out in Paper~I, we first carry out an exploratory Bayesian analysis to constrain the relevant chameleon parameter space. We recall that, for our purposes, the full solar chameleon parameter space is five-dimensional, and described by the parameter vector $\boldsymbol{\theta} \equiv \{\beta_e,\beta_{\gamma},M_e,\Lambda, n\}$, where $\beta_e$ and $M_e$ are the electron conformal coupling and disformal scale respectively, $\beta_{\gamma}$ is the photon conformal coupling, and $\Lambda$ and $n$ are the energy scale and power-law index describing the chameleon self-interaction potential. The data vector we consider, which we denote by $\boldsymbol{d}$, consists of binned measurements of the differential electron recoil event rate per unit electron recoil energy, within bins of size $\Delta E_R=1\,{\text{keV}}$. We then model the likelihood to observe the data $\boldsymbol{d}$ given the model parameters $\boldsymbol{\theta}$, ${\cal L}(\boldsymbol{\theta} \vert \boldsymbol{d})$, as follows:
\begin{equation}
{\cal L}(\boldsymbol{\theta} \vert \boldsymbol{d}) = \exp \left ( -\frac{\chi^2(\boldsymbol{\theta},\boldsymbol{d})}{2} \right ) \,,
\label{eq:likelihood}
\end{equation}
where we have defined the $\chi^2$ function as follows:
\begin{eqnarray}
\chi^2(\boldsymbol{\theta},\boldsymbol{d})=\sum _i \left [ \frac{d_i- \left ( B_0+ \left ( \dfrac{{\rm d}R}{{\rm d}E_R} \right ) _{\text{th}}(\boldsymbol{\theta}) \right ) }{\sigma_i} \right ] ^2\,,
\label{eq:chi2}
\end{eqnarray}
where the sum extends over all the bins in $E_R$, with $d_i$ and $\sigma_i$ being the $i$th event rate measurement and uncertainty respectively (see the black data points in Fig.~5 of Ref.~\cite{XENON:2022ltv}), $({\rm d}R/{\rm d}E_R)_{\text{th}}$ denotes the theoretical event rate calculated using Eq.~(\ref{eq:drder}), and $B_0$ is the energy-dependent background model (see the red curve in Fig.~5 of Ref.~\cite{XENON:2022ltv}).~\footnote{Both the binned event rates and the background have been taken from the publicly available repository \href{https://zenodo.org/records/7311940}{zenodo.org/records/7311940}.} Following the indications of the XENONnT collaboration from Ref.~\cite{XENON:2022ltv}, we consider an energy threshold of $E_R=1\,{\text{keV}}$, and therefore event rate measurements from $1\,{\text{keV}}$ up to $30\,{\text{keV}}$.

In our initial Bayesian analysis, we explore the full five-dimensional parameter space characterized by $\boldsymbol{\theta}$. With the exception of $n$, these parameters span several orders of magnitude, and we therefore adopt priors which are flat in the logarithm thereof, as commonly done in these situations. Specifically, we set uniform priors on $\log_{10}\beta_e \in [0,2]$, $\log_{10}\beta_{\gamma} \in [0,15]$, $\log_{10}M_e \in [0,6]$, $\log_{10}(\Lambda/{\text{eV}}) \in [-8,0]$, and $n \in [0,6]$. We stress that most of these priors are deliberately large and generous, allowing for regions of parameter space which are in principle excluded by independent experiment. For instance, the CAST helioscope sets the upper limit $\beta_{\gamma} \lesssim 5.7 \times 10^{10}$~\cite{CAST:2018bce}, whereas constraints from the Large Electron-Positron Collider (LEP) and the Large Hadron Collider (LHC) require $M_e \gtrsim 3\,{\text{GeV}}$~\cite{Brax:2014vva} and $M_e \gtrsim 650\,{\text{GeV}}$~\cite{Brax:2015hma} respectively (although with important caveats we will discuss later). The reason is that our first objective is to set completely independent constraints on these parameters from XENONnT alone, and only later compare these constraints to those from independent observations. We sample the posterior distributions of these parameters via Markov Chain Monte Carlo (MCMC) methods, through the affine-invariant ensemble sampler \texttt{emcee}~\cite{Foreman-Mackey:2012any}. The resulting MCMC chains are then analyzed using the \texttt{GetDist} package~\cite{Lewis:2019xzd}.

While a Bayesian analysis of the full five-dimensional parameter space relevant to the problem provides the most direct comparison to Paper~I, it is not necessarily the most physically meaningful approach in our case, for two main reasons. Firstly, as already observed in Paper~I, most of the parameters are largely unconstrained, and the theoretical signal ends up mostly depending on a specific combination of $\beta_{\gamma}$ and $M_e$, which we refer to as $\beta_{\text{eff}}$: this combination is proportional to $\beta_{\gamma}M_e^{-4}$, and the physical reason for this scaling will be discussed in Sec.~\ref{sec:resultsdiscussion}, although we anticipate that it is analogous to $g_{a\gamma}g_{ae}$ in the solar axion case. Moreover, the absence of an excess means that only upper limits can be obtained: for instance, in Paper~I fitting the excess required a specific range of $\beta_{\text{eff}}$ parameter space, at face value excluded by independent probes, whereas in our case we can expect to obtain an upper limits on $\beta_{\text{eff}}$. In such a scenario, marginalizing over largely unconstrained directions will add little physical information.

For this reason, in a second stage we adopt a frequentist, profile-likelihood approach aimed at setting upper limits. In direct analogy with the standard mass versus cross-section/coupling choice of variables for weakly interacting massive particles (WIMPs), light DM, axions, and axion-like particles, we show our results in the $\beta_e$-$\beta_{\text{eff}}$ plane, with $\Lambda=2.4\,{\text{meV}}$ fixed to the DE scale, while also fixing $n=1$. In our case, once $\Lambda$ and $n$ are fixed, $\beta_e$ determines the effective chameleon mass (and hence the degree of screening), whereas $\beta_{\text{eff}}$ controls the overall detection rate. This choice of variables allows for a transparent comparison with external bounds, since $\beta_e$ controls the coupling of the chameleon to matter and is therefore the parameter most directly constrained by laboratory, astrophysical, and fifth-force experiments. For $n=1$ chameleons, atom interferometry and torsion balance searches leave open the window in parameter space where $20 \lesssim \beta_e \lesssim 300$~\cite{Burrage:2017qrf}, although the $5 \lesssim \beta_e \lesssim 630$ region was recently excluded by levitated force sensors~\cite{Yin:2022geb}. At any rate, following the same logic as earlier, we will work within a deliberately larger range of $\beta_e$ values, in the interest of obtaining completely independent limits from XENONnT alone.

Operationally speaking, we proceed as follows. We consider fixed values of $\log_{10}\beta_e \in [0,4]$. For each of these fixed values $\widetilde{\beta}_e$, we vary $\beta_{\text{eff}}$ and determine the value of $\beta_{\text{eff}}$ that minimizes the $\chi^2$ in Eq.~(\ref{eq:chi2}). We refer to this value as $\widetilde{\beta}_{\text{eff}}$, and the corresponding $\chi^2$ to $\widetilde{\chi}^2_{\min}$, where the tilde highlights that this is done for the given value of $\widetilde{\beta}_e$. Following Wilks' theorem, we determine the one-sided 95\% confidence level (C.L.) upper limit on $\beta_{\text{eff}}$, which we refer to as $\widetilde{\beta}_{\text{eff},95}$, by the condition $\chi^2(\widetilde{\beta}_{\text{eff},95},\widetilde{\beta}_e)=\widetilde{\chi}^2_{\min}+2.71$, appropriate for a one-sided limit on a single parameter of interest~\cite{Wilks:1938dza}. Our main result are therefore frequentist upper limits on $\beta_{\text{eff}}$ as a function of $\beta_e$. As we shall see, these limits are essentially flat along the $\beta_e$ direction, and the corresponding 95\%~C.L. upper limit on the effective coupling is $\beta_{\text{eff}}<10^{-6.9}$.

\section{Results and discussion}
\label{sec:resultsdiscussion}

\begin{figure}[!htbp]
\centering
\includegraphics[width=0.9\linewidth]{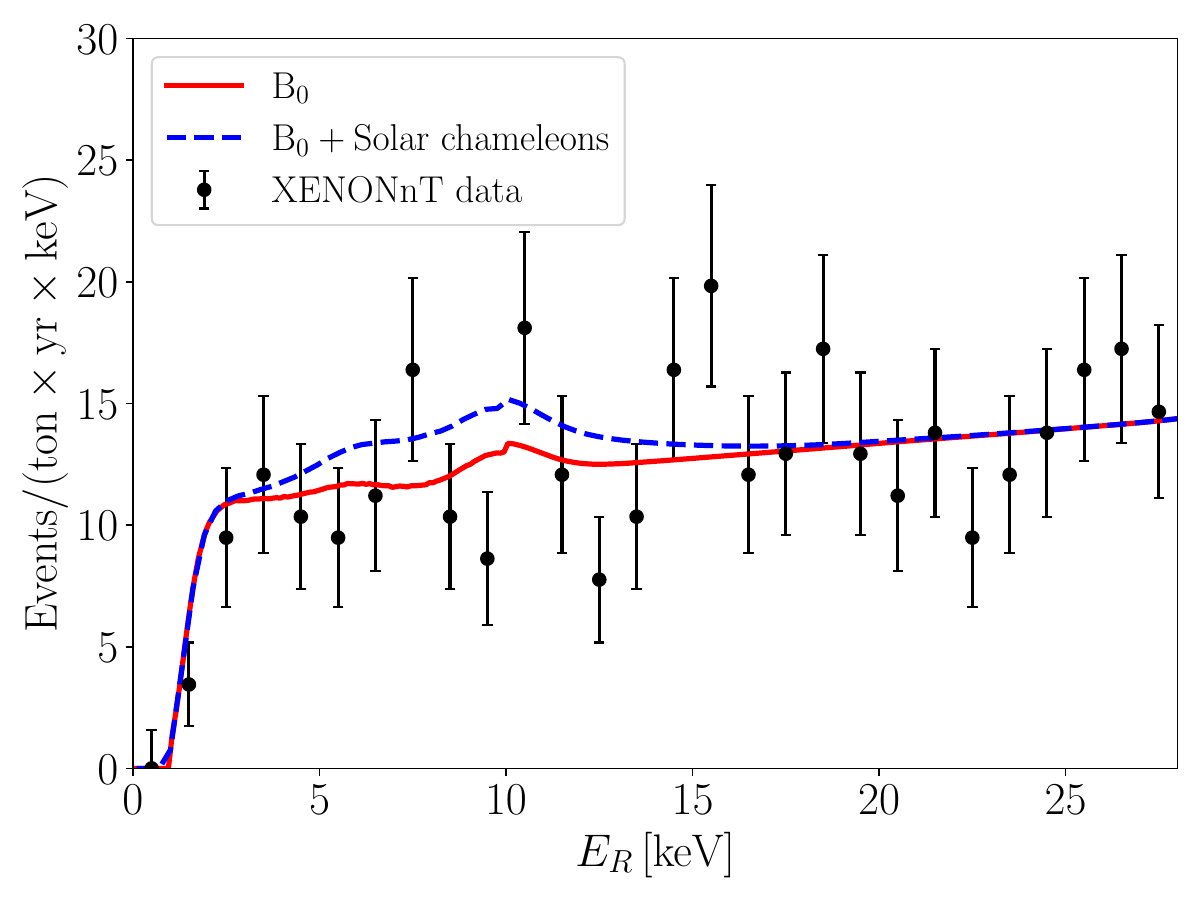}
\caption{Benchmark example of solar chameleon fit to the XENONnT electron recoil event rate, with chameleon parameters fixed to $\beta_e=10$, $\beta_{\gamma}=10^{9.4}$, $M_e=10^4\,{\text{eV}}$, $\Lambda=2.4\,{\text{meV}}$, and $n=1$. This choice of parameters corresponds to an effective coupling strength $\beta_{\text{eff}}=10^{-6.6}$ [see Eq.~(\ref{eq:betaeff})], which is allowed by the data (see Fig.~\ref{fig:betaeffupperlimit}). The XENONnT data is given by the black datapoints, the solid red curve is the XENONnT energy-dependent background $B_0$, and the dashed blue curve is the total (background plus signal) solar chameleon event rate. Detector efficiency and energy resolution effects are already accounted for when plotting the total signal.}
\label{fig:benchmarksignal}
\end{figure}

We begin by noting that there are benchmark points in parameter space which can slightly improve the fit relative to the background-only hypothesis, though at a level far from sufficient to claim a detection of solar chameleons. These serve as useful reference points for understanding the qualitative features of the response of the XENONnT electron recoil signal to the solar chameleon parameters. Since XENONnT electron recoil data show no evidence for an excess above the background $B_0$, we do not aim to identify best-fit regions or perform a detailed parameter estimation: in fact, because of the absence of a clear signal, we do not expect that further quantification of local best fits or confidence regions would provide additional physical insight. One such benchmark is shown in Fig.~\ref{fig:benchmarksignal}, where we set $\beta_e=10$, $\beta_{\gamma}=10^{9.4}$, $M_e=10^4\,{\text{eV}}$, $\Lambda=2.4\,{\text{meV}}$ (at the DE scale), and $n=1$: from Fig.~\ref{fig:chameleonescape} we see that this benchmark is physically interesting, i.e.\ that the chameleons can escape from the solar core. We see that at this benchmark point the chameleon contribution slightly improves the fit to the signal at $E_R=7\,{\text{keV}}$ and $10\,{\text{keV}}$, which nevertheless lie only $\gtrsim 1\sigma$ above the background. We stress that the benchmark point in parameter space we have considered is reportedly excluded by levitated force sensor constraints, which closed the remaining window in $\beta_m$ parameter space for $n=1$ chameleons~\cite{Yin:2022geb}, as well as LHC constraints on $M_e$~\cite{Brax:2015hma}. Nevertheless, as we will also discuss in much more detail later, the levitated force sensor constraints do not apply to $n \neq 1$, and $n$ essentially has no impact on the XENONnT signal (so we might as well have considered a different $n$ from the start, although we have stuck to $n=1$ as this is by far the most studied case), whereas the LHC constraints do not directly apply to our scenario, as also discussed in Paper~I for LEP. In any case, given the absence of a statistically significant signal, we turn to setting exclusion limits on the chameleon parameter space.

We begin by carrying out an exploratory Bayesian analysis, mirroring what was done in Paper~I. As anticipated, and as observed in Paper~I, most of the parameters remain unconstrained. For this reason we relegate the joint posteriors of the five parameters, shown in Fig.~\ref{fig:posteriors}, to Appendix~\ref{app:posteriors}. We have checked that for the three parameters which remain unconstrained, i.e.\ $\beta_e$, $\Lambda$, and $n$, it is necessary to move to unphysical values far beyond the already generous prior ranges before any appreciable impact on the event rate becomes visible. This is particularly true for $n$, as we have observed that one has to go to $n \gtrsim 200$ before there is even a minimal effect on the predicted event rate.~\footnote{While such steep Ratra–Peebles potentials are formally allowed, they are usually considered to be theoretically unmotivated, as they imply an excessively rapid density dependence of the effective mass and lie well outside the regime where the chameleon mechanism is under theoretical control.}

A different behaviour is observed for $\beta_{\gamma}$ and $M_e$, which we find to be strongly degenerate. The degeneracy describes an anti-correlation between the two, i.e.\ $\beta_{\gamma}$ increases as $M_e$ decreases, and viceversa. The reason can be understood as follows: since energy resolution and detection efficiency effects do not alter our arguments, let us consider the raw event rate given in Eq.~(\ref{eq:dr0der}). The flux term, i.e.\ ${\rm d}\dot{N}_{P}/{\rm d}E_R+{\rm d}\dot{N}_{B}/{\rm d}E_R$, scales as $\beta_{\gamma}^2$, for reasons discussed at the end of Sec.~\ref{subsubsec:primakoff}, and as can be seen in Eqs.~(\ref{eq:productionprimakoff},\ref{eq:productionmagnetic}). We now turn to the cross-section term $\sigma_{\phi e}$, which is given by Eq.~(\ref{eq:sigmaphie}). Let us assume for the moment that the disformal term dominates, i.e.\ $\sigma_{\phi e} \approx m_e^2E_R^4/8\pi^2M_e^8$. Then, the differential event rate ${\rm d}R/{\rm d}E_R$ scales as $\beta_{\gamma}^2M_e^{-8}$, i.e.\ as $\beta_{\text{eff}}^2$, where we define the effective coupling strength $\beta_{\text{eff}}$ as follows:
\begin{equation}
\beta_{\rm eff} \equiv \beta_{\gamma} \left ( \frac{\text{eV}}{M_e} \right ) ^4\,.
\label{eq:betaeff}
\end{equation}
Note that the above expression is manifestly dimensionless.~\footnote{Note that Paper~I adopts a slightly different normalization for $M_e$, with ${\text{keV}}$ instead of ${\text{eV}}$ in Eq.~(\ref{eq:betaeff}): the two definitions therefore differ by an additive factor of $12$.} The $\beta_{\gamma}$-$M_e$ degeneracy direction observed in our posteriors is indeed precisely described by Eq.~(\ref{eq:betaeff}), i.e.\ it corresponds to combinations of $\beta_{\gamma}$ and $M_e$ which maintain $\beta_{\text{eff}}$ constant.

The question is then why the disformal contribution dominates the total cross-section. This was indeed already observed in Paper~I, but not explained in detail at the time. The explanation is actually rather simple, and is easily understood by inspecting the relevant Lagrangian terms, i.e.\ the first and third terms in the second row of Eq.~(\ref{eq:effectivefinal}). The disformal coupling to electrons is described by a dimension-8 operator (which is indeed suppressed by $M_e^4$), so one would na\"{i}vely expect it to be the subdominant one. The key point, however, is that the operator describing the conformal coupling, despite being a dimension-5 one, is a \textit{Planck-suppressed} operator. In fact, the relevant energy scale for the conformal coupling is $M_{\text{Pl}}/\beta_e$. Unless $\beta_e$ is several orders of magnitude larger than unity, which is excluded at high statistical significant by atom interferometry and torsion balance searches~\cite{Burrage:2017qrf} all (scattering/absorption) processes associated to this coupling will be significantly suppressed, explaining why the conformal piece of $\sigma_{\phi e}$ is swamped by the disformal one, whose energy scale $M_e$ can in principle be much lower (more on this later). Therefore, the relevant coupling for the detection of solar chameleons in direct detection experiments is the disformal one.

\begin{figure}[!htbp]
\centering
\includegraphics[width=0.9\linewidth]{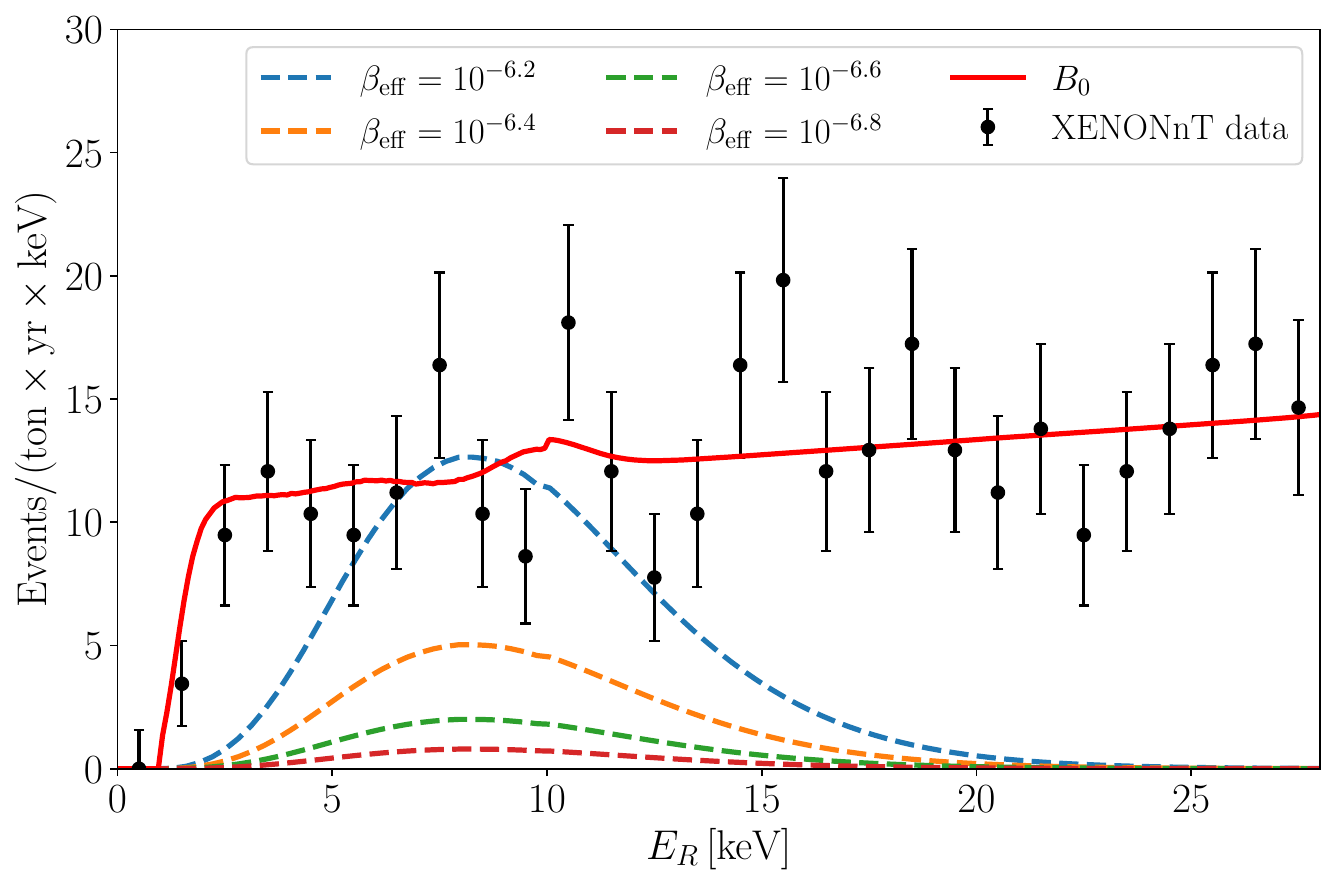}
\caption{As in Fig.~\ref{fig:benchmarksignal}, but isolating the solar chameleon contribution to the total signal, for different values of the effective coupling strength $\beta_{\text{eff}}$ between $10^{-6.8}$ and $10^{-6.2}$, as indicated by the color coding. Note that the solar chameleon contribution needs to be summed to the energy-dependent background $B_0$ in order to be meaningfully compared to the XENONnT electron recoil data. It is visually clear that, for most of the plotted values of $\beta_{\text{eff}}$, the total (chameleon+background) signal would exceed the XENONnT measurements, and the frequentist limits we will later obtain on $\beta_{\text{eff}}$ should therefore fall broadly within the plotted range.}
\label{fig:eventsbetaeff}
\end{figure}

\begin{figure}[!htbp]
\centering
\includegraphics[width=0.9\linewidth]{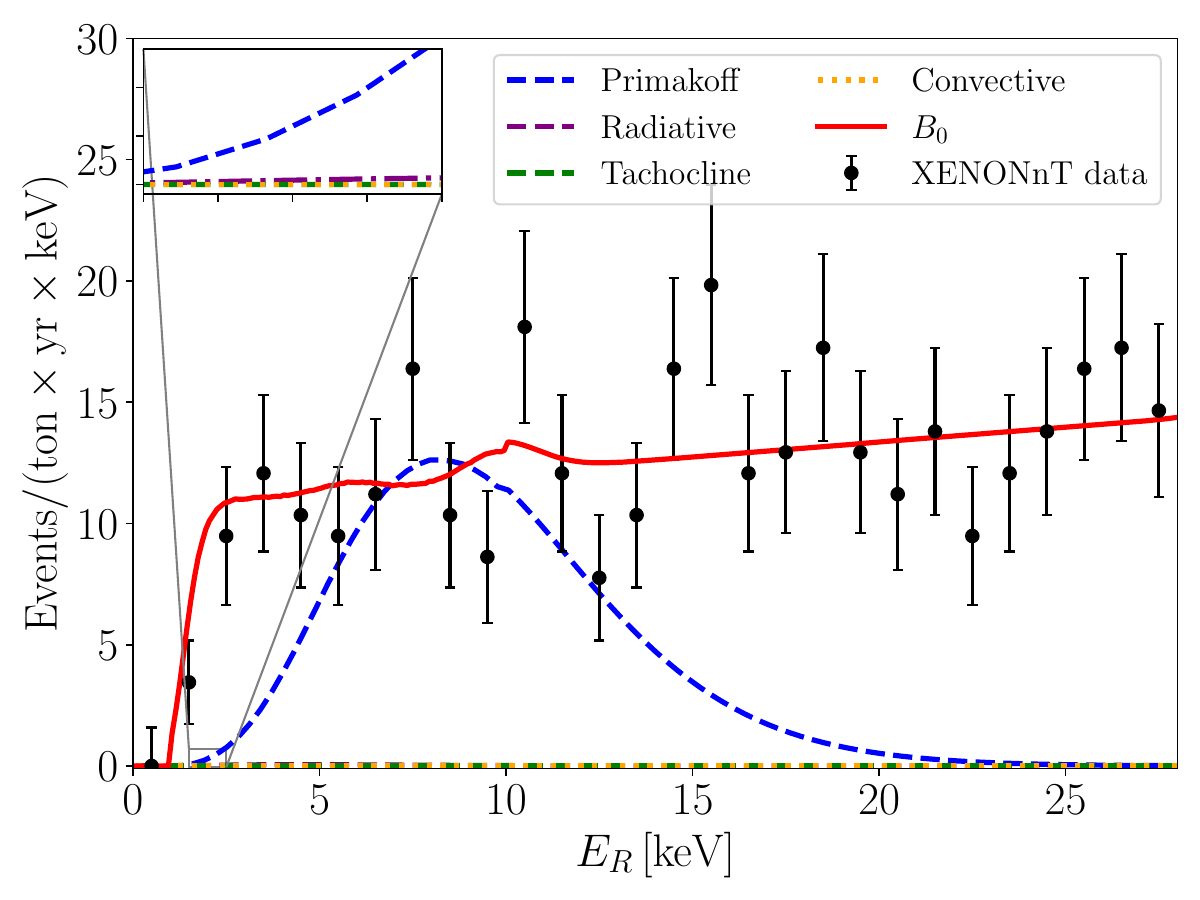}
\caption{As in Fig.~\ref{fig:benchmarksignal}, but for an effective coupling strength $\beta_{\text{eff}}=10^{-6.2}$, and separating the contributions to the chameleon signal arising from different components of the solar chameleon spectrum: Primakoff production from transverse photons in the electric fields of electrons and ions (dashed blue curve), and magnetic conversion from the solar bulk magnetic field within the radiative zone (dashed violet curve), tachocline (dashed green curve), and convective zone (dotted yellow curve). Of all the components, the Primakoff one clearly dominates by a large margin, given the recoil energy range of interest to XENONnT (compare against the spectrum of Fig.~\ref{fig:spectrum}). The inset zooms into the low-energy region where the relative contribution from the bulk magnetic field components is least insignificant. Note that the chosen value of $\beta_{\text{eff}}$ is excluded by the data (see Fig.~\ref{fig:betaeffupperlimit}; note that the four components need to be summed to the energy-dependent background $B_0$ in order to be meaningfully compared to the XENONnT electron recoil measurements), and has been chosen to grossly exaggerate the difference between the components, since the bulk magnetic field ones would otherwise be impossible to see by eye.}
\label{fig:signalcomponents}
\end{figure}

Detection of solar chameleons is controlled by the effective coupling $\beta_{\text{eff}}^2$, which includes the combined effect of production in the Sun (spectrum $\propto \beta_{\gamma}^2$) and detection on Earth (cross-section $\propto M_e^{-8}$). As anticipated earlier, precisely the same thing happens with solar axions, where it is instead the product of the axion-photon and axion-electron couplings $g_{a\gamma}g_{ae}$ which plays a key role, for exactly the same reasons: the former controls production in the Sun, whereas the latter controls detection on Earth (see e.g.\ Ref.~\cite{XENON:2020rca}). In Fig.~\ref{fig:eventsbetaeff}, considering $n=1$ chameleons while fixing $\Lambda=2.4\,{\text{meV}}$ and $\beta_e=10$, we plot the chameleon contribution to the event rate for representative values of $\log_{10}\beta_{\text{eff}}$ between $-6.8$ and $-6.2$ (note that these contributions should be summed to the background $B_0$ in order to be compared with the data). Already by eye we can easily appreciate that, for sufficiently large values of $\beta_{\text{eff}}$ within this range, the total (signal plus background) predicted event rate will exceed the measured data. Therefore, we anticipate that the frequentist limits we will later obtain on $\beta_{\text{eff}}$ should lie broadly within the range plotted in Fig.~\ref{fig:eventsbetaeff}. Focusing instead on an effective coupling strength $\beta_{\text{eff}}=10^{-6.2}$, in Fig.~\ref{fig:signalcomponents} we plot the contributions to the chameleon signal arising from different components of the solar chameleon spectrum: Primakoff production from transverse photons in the electric fields of electrons and ions, and magnetic conversion within the radiative zone, tachocline, and convective zone. Although such a value of $\beta_{\text{eff}}$ will be excluded a posteriori, we have chosen it in order to grossly exaggerate the differences between the components shown, as the magnetic conversion ones would otherwise be impossible to see by eye. It is clear that the component arising from Primakoff production strongly dominates at all energies. Moreover, the contribution peaks at larger energies compared to those of the magnetic conversion components. This highlights the importance, especially at the energies of interest, of adopting the updated solar chameleon spectrum studied in Paper~II, given that the (dominant) Primakoff component had been completely neglected in the earlier analysis of Paper~I.

\begin{figure}[!htbp]
\centering
\includegraphics[width=0.9\linewidth]{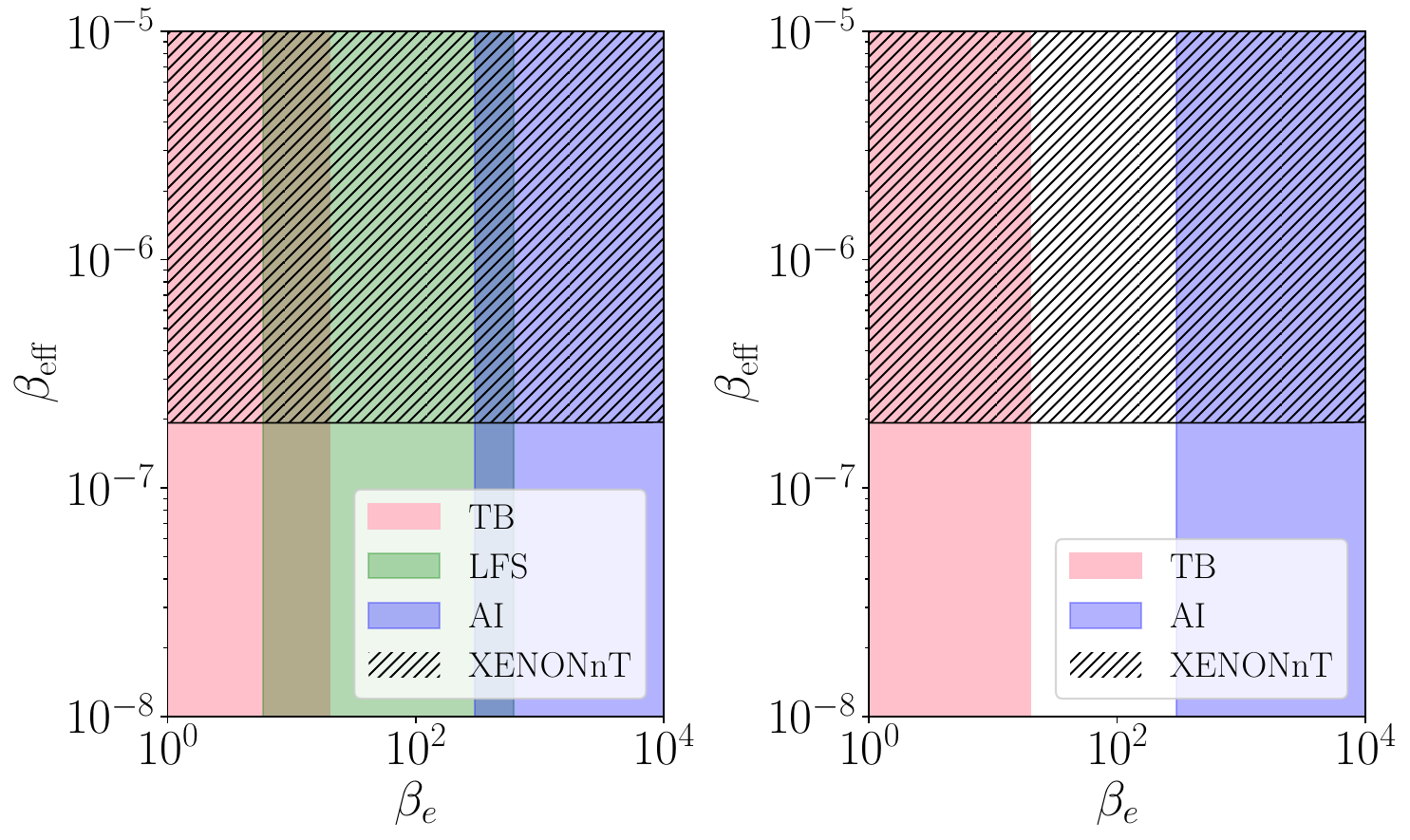}
\caption{\textit{Left panel}: frequentist 95\%~C.L. upper limit on the effective coupling strength $\beta_{\text{eff}}$ as a function of the conformal coupling to electrons $\beta_e$, obtained from XENONnT electron recoil data, for $n=1$ chameleons. This limit is to good approximation independent of $\beta_e$, and indicates $\log_{10}\beta_{\text{eff}}<-6.9$. The region excluded by this limit is that hatched in black. The vertical shaded bands indicate the regions excluded by atom interferometry (AI, blue region), levitated force sensors (LFS, green region), and torsion balance (TB, red region) searches. \textit{Right panel}: same as in the left panel, but for the case of $n=4$ chameleons, where the LFS constraints do not apply.}
\label{fig:betaeffupperlimit}
\end{figure}

We now compute our frequentist, profile-likelihood upper limits on the effective coupling which we report in the $\beta_e$-$\beta_{\text{eff}}$ plane. We recall that this is in analogy to the mass versus cross-section choice for constraints resulting from DM searches. In addition, recall that we fix $\Lambda=2.4\,{\text{meV}}$ and $n=1$, while working within the very generous range of conformal coupling $\log_{10}\beta_e \in [0,4]$. The result is shown in the left panel of Fig.~\ref{fig:betaeffupperlimit}, where we also show the regions excluded by torsion balance (TB), atom interferometry (AI), and levitated force sensor (LFS) searches, with the latter constraining the remaining $\beta_m$ window for $n=1$ chameleons. We see that the 95\%~C.L upper limit on $\beta_{\text{eff}}$ is essentially flat, i.e.\ independent of $\beta_e$. In particular, we find the 95\%~C.L. upper limit $\log_{10}\beta_{\text{eff}}<-6.9$.

The upper limit $\log_{10}\beta_{\text{eff}}<-6.9$ does not directly distinguish between production and detection effects. However, if independent limits exist on either the chameleon-photon coupling $\beta_{\gamma}$ or the scale of the disformal coupling to electrons $M_e$, this limit can be readily recast as a bound on the other quantity. For instance, the tightest upper limit on $\beta_{\gamma}$ comes from the CAST helioscope, which sets $\beta_{\gamma} \lesssim 5.7 \times 10^{10}$~\cite{CAST:2018bce}, i.e.\ $\log_{10}\beta_{\gamma} \lesssim 10.8$. Folding this into our $\beta_{\text{eff}}$ limit we see that it translates into a lower limit on $4\log_{10}(M_e/{\text{eV}}) \gtrsim 17.7$, i.e.\ $M_e \gtrsim 25.1\,{\text{keV}}$. We note that this upper limit is still consistent with the benchmark value of $M_e \approx 4\,{\text{MeV}}$ required to fit the XENON1T excess with solar chameleons, identified in Paper~I. Conversely, the tightest constraints on the disformal coupling to electrons come from the LHC, which require $M_e \gtrsim 650\,{\text{GeV}}$~\cite{Brax:2015hma}. Folding this limit into our $\beta_{\text{eff}}$ constraint leads to an upper limit on $\log_{10}\beta_{\gamma} \lesssim 40.3$, i.e.\ $\beta_{\gamma} \lesssim 2.0 \times 10^{40}$. Therefore, the constraints we obtain are $\beta_{\gamma} \lesssim 2.0 \times 10^{40}$ (XENONnT+LHC) and $M_e \gtrsim 25.1\,{\text{keV}}$ (XENONnT+CAST): both limits are rather weak and therefore not competitive with independent existing bounds -- nevertheless, as we will stress shortly, our limits can be extended essentially to arbitrary $n$, which makes them considerably more general and valuable than they may appear at first glance.

Before moving on, two comments are in order concerning the CAST and LHC limits on $\beta_{\gamma}$ and $M_e$ respectively. The upper limit on $\beta_{\gamma}$ has been obtained by only considering a simplistic model for chameleon production limited to the tachocline (the same model considered in Paper~I). Adopting the full spectrum we used here and demanding that chameleons do not carry more than 3\% of the solar luminosity, as inferred from recent global fits to helioseismic and solar neutrino observables~\cite{Vinyoles:2015aba}, already results in a tighter limit $\beta_{\gamma} \lesssim 10^{10}$, as explained in Paper~II. Still in Paper~II, we provided an approximate, order-of-magnitude estimate of how much the CAST bound would tighten if the full chameleon spectrum were taken into account, finding $\beta_{\gamma} \lesssim 6 \times 10^8$. Therefore, the quoted limit $\beta_{\gamma} \lesssim 5.7 \times 10^{10}$ should be regarded as a somewhat overly conservative one. The converse is true for the limit $M_e \gtrsim 3\,{\text{GeV}}$. As stressed in Paper~I, this limit was obtained assuming a massless chameleon and therefore should not directly apply to our scenario. However, in order to be reliably applied, this limit should be properly revised by determining the chameleon profile in the LHC pipe simultaneously including conformal and disformal couplings (see e.g.\ Ref.~\cite{Argyropoulos:2023pmy}). Therefore, one may regard our comparatively weak lower limit on $M_e$ as a fully independent and valuable constraint, obtained without invoking the assumptions underlying collider-based analyses.

As further confirmation that this limit is essentially insensitive to $n$, we recompute it for the interesting benchmark case $n=4$, here taken as a representative steeper inverse-power case which has been studied in the literature (including in Paper~II), while once more fixing $\Lambda=2.4\,{\text{meV}}$. The result is shown in the right panel of Fig.~\ref{fig:betaeffupperlimit}. We see that the resulting 95\%~C.L upper limit on $\beta_{\text{eff}}<-6.9$ is once more independent of $\beta_e$, and essentially identical to that obtained in the $n=1$ case. In fact, as stressed earlier, only for theoretically unmotivated large values $n \gtrsim 200$ is there a rather minimal effect on the predicted event rate. Therefore, we can conclude that our limit $\beta_{\text{eff}}$ \textit{is virtually independent of the chameleon power-law index} $n$. This result is particularly significant because most existing constraints on chameleons in the literature only apply to selected values of $n$. In the vast majority of cases, constraints are reported for the $n=1$ case, while in a few others they are extended to $n=4$, $n=-4$, or in any case relatively small values of $n$: as an example, Fig.~6 of Ref.~\cite{Burrage:2017qrf} reaches $n=13$, but we are not aware of other studies extending to larger $n$. For instance, the CAST upper limit on $\beta_{\gamma}$ has been explicitly computed for $n=1$~\cite{CAST:2018bce}. Similarly, the LFS search that excludes the remaining allowed region for $\beta_e$ does so under the assumption of $n=1$ chameleons~\cite{Yin:2022geb}. While at first glance weak, our limit on $\beta_{\text{eff}}$ applies broadly across the entire class of inverse power-law chameleon models of theoretical interest, rather than for a single benchmark case, making it one of the most generally applicable constraints on chameleons to date. It is worth stressing once more that our limit is obtained in the theoretically strongly motivated case where $\Lambda$ is fixed to the DE scale.

An important caveat to all our results, as stressed earlier, is that the production mechanisms presented in Sec.~\ref{subsubsec:primakoff} and~\ref{subsubsec:magnetic} do not account for the contributions of longitudinal plasmons to the solar chameleon spectrum. However, a preliminary estimate suggests that the contribution of longitudinal plasmons is subdominant in the energy range relevant to XENONnT and may be safely neglected at this stage, although a comprehensive, quantitative treatment of the plasmon contribution is among our plans for a separate follow-up work to Paper~II. At any rate, including additional plasmon-related channels would only increase the predicted chameleon flux, thereby strengthening the expected signal in the XENONnT detector. Our limit on $\beta_{\text{eff}}$ should thus be regarded as conservative, since the adopted theoretical prediction for the signal represents only a lower limit to the true expectation. Including the contribution from longitudinal plasmons can therefore only tighten the upper limit on $\beta_{\text{eff}}$.


We conclude with a brief, more speculative discussion on how solar chameleons could be distinguished from solar axions, highlighting qualitative differences that may be relevant for future searches such as the one considered here. In the case of solar axions, transverse photons dominate the production via the Primakoff process, in which photons are converted into axions in the electric fields of electrons and ions. Longitudinal modes do not contribute to axion production in the absence of magnetic fields. However, if magnetic fields exist in the solar core (as assumed in this work) axions can also be produced through the conversion of longitudinal photons in those fields, with distinctive spectral features discussed e.g.\ in Refs.~\cite{Guarini:2020hps,Caputo:2020quz}. Scalar particles such as chameleons, on the other hand, can be generated through the interaction of longitudinal plasmons even in the absence of a core magnetic field, directly in the electric fields of electrons and ions. Therefore, both solar chameleons and solar axions can in principle receive contributions from longitudinal modes, though through rather distinct physical mechanisms. A careful comparison between the solar chameleon and axion spectra could therefore, with sufficient statistics, allow one to distinguish the two in the case of a positive experimental signal. However, the similarities between the two spectra, together with the still uncertain details of the chameleon spectrum generated by longitudinal plasmons, make this strategy challenging. These issues are particularly relevant for experiments sensitive to sub-${\text{keV}}$ spectra, such as IAXO~\cite{IAXO:2019mpb,IAXO:2024wss,IAXO:2020wwp}. }

Multi-tonne DM direct detection experiments, such as XENONnT, can also access recoil energies lower than the ones considered here. In particular, their lower effective energy thresholds make analyses based solely on the ionization signal particularly sensitive to sub-${\text{keV}}$ electron recoils resulting from interactions with low-energy solar chameleons. We recall that the XENONnT analysis considered here requires coincident detection of both the primary scintillation (S1) and secondary ionization (S2) signals, with the resulting energy threshold being mostly set by the S1 requirement. However, we note that S2-only analyses have also been carried out within XENONnT~\cite{XENON:2024znc} (as well as XENON1T~\cite{XENON:2019gfn} and PandaX~\cite{PandaX:2022xqx}), and are sensitive to recoil energies an order of magnitude lower than those relevant to this work, albeit at the price of a lower signal-to-noise ratio.

While their spectral features may be similar, a more fundamental qualitative difference between scalars and pseudoscalars emerges in their detection channels, particularly in large underground experiments such as XENONnT. As we have seen, chameleons are detected primarily through the disformal contribution to the cross-section, whereas axion detection proceeds via the axio-electric effect. Aside from the different energy dependence of the two cross-sections (which can already serve as a discriminant, given sufficient energy resolution) the axio-electric cross-section scales roughly as $Z^5$ with the atomic number of the target atoms, reflecting its proportionality to the photoelectric cross-section. The disformal cross-section, on the other hand, does not inherit this dependence and is nearly target-independent. Therefore, in the event of a positive signal in direct detection experiments, a multi-target confirmation (for instance using detectors based on xenon, argon, germanium, or silicon) would provide a crucial test for a chameleon versus axion origin. In summary, aside from different spectral scalings due to the distinct energy dependence of the relevant cross-sections, a genuine solar chameleon signal should be seen (up to trivial electron density per unit mass and threshold/efficiency effects) with comparable strength across detectors based on different target materials, whereas an axion signal would be expected to vary strongly with target material, while at the same time exhibiting target-specific features.

\section{Conclusions}
\label{sec:conclusions}

New (ultra)light scalar particles are not only theoretically well-motivated, but may also underlie the dark matter and dark energy phenomena. In the latter case, however, the key challenge is to reconcile the existence of an ultralight scalar field driving cosmic acceleration with stringent constraints on the existence of fifth forces. Chameleon scalars provide a concrete example of how such forces can be dynamically screened, allowing the scalar field to remain active on cosmological scales while making the associated fifth force extremely short-ranged (and thus below current experimental sensitivity) in high-density environments. This mechanism operates via a direct coupling to matter fields, which in turn inevitably makes chameleons amenable to laboratory searches. In fact, focusing on electron recoil data from the XENON1T experiment (which reported indications of a low-energy excess), earlier proof-of-principle work by three of us in Ref.~\cite{Vagnozzi:2021quy} demonstrated that chameleon particles produced within the Sun can be detected in terrestrial dark matter direct detection experiments, opening a new window onto the physics of dark energy. Since then, significant progress has been made both on the theoretical side, with substantial improvements in the modeling of solar chameleon production by five of us in Ref.~\cite{OShea:2024jjw}, and on the experimental side, as XENONnT has become operational, with a larger exposure and substantially reduced background compared to XENON1T, and no indication of excesses~\cite{XENON:2022ltv}. Motivated by these significant developments, in the present work we have reassessed prospects for direct detection of solar chameleons, focusing on XENONnT electron recoil data, and setting novel limits on the chameleon model parameters.

Building on the improved modeling of the solar chameleon spectrum from Ref.~\cite{OShea:2024jjw}, we find that the dominant contribution to the electron recoil signal originates from the Primakoff component of the solar chameleon spectrum, produced in the electric fields of electrons and ions in the solar plasma (see Fig.~\ref{fig:signalcomponents}). This leads to a substantially larger predicted event rate compared to the previous estimate of Ref.~\cite{Vagnozzi:2021quy}, which only included the contribution from magnetic conversion in the tachocline, while also resulting in a signal peaking at higher energies. Consistently with previous analyses, we show that XENONnT data are unable to set meaningful constraints on the conformal coupling to electrons, $\beta_e$, as well as the parameters $\Lambda$ and $n$ characterizing the chameleon cosmological dynamics. Instead, we find that the signal is controlled by a single effective coupling strength $\beta_{\text{eff}} \propto \beta_{\gamma}M_e^{-4}$, which captures the combined effects of production in the Sun (spectrum $\propto \beta_{\gamma}^2$) and detection on Earth (cross-section $\propto M_e^{-8}$, since the conformal contribution to the detection cross-section is Planck-suppressed), with $\beta_{\gamma}$ and $M_e$ the chameleon-photon coupling and scale of the disformal coupling to electrons respectively. When fixing $\Lambda$ to the dark energy scale, $\Lambda=2.4\,{\text{meV}}$, we find the upper limit $\log_{10}\beta_{\text{eff}}<-6.9$. While for $n=1$ chameleons this bound does not translate into particularly stringent limits on $\beta_{\gamma}$ or $M_e$ once external constraints such as those from CAST or LHC are taken into account, the result remains very significant as it is virtually independent of both $\beta_e$ and especially the index $n$. Therefore, our limit on $\beta_{\text{eff}}$ applies broadly across the entire class of inverse power-law chameleon models of theoretical interest, extending well beyond the $n=1$ case that is almost exclusively considered in existing analyses, and reportedly excluded by levitated force sensor experiments: this makes our result one of the most generally applicable constraints on chameleons to date.

Overall, our results demonstrate that existing multi-tonne direct detection experiments such as XENONnT can already place meaningful constraints on chameleon particles: this substantially broadens the scientific scope of terrestrial dark matter experiments, which can also probe screened scalars at no additional experimental cost, using data collected for ``standard'' WIMP or axion searches. Our study also points to several interesting directions for future work. A first key step, already emphasized in Ref.~\cite{OShea:2024jjw}, is to include the longitudinal plasmon contribution to the solar chameleon flux, expected to be most relevant at low energies. The resulting low-energy solar chameleon spectrum would allow targeted searches in helioscopes such as IAXO~\cite{IAXO:2019mpb}, or in ionization (S2)-only analyses within experiments such as XENONnT~\cite{XENON:2024znc}. It would also be interesting to extend our analysis to other screened dark energy models, such as symmetrons. Since most screening mechanisms (including the chameleon one) can be interpreted as a scalar-tensor modification of gravity in the Jordan frame, our results may be viewed as a step towards direct detection of scalar degrees of freedom arising in theories of modified gravity. In addition, it could be particularly interesting to connect to recent cosmological hints for a dynamical dark energy component~\cite{DESI:2024mwx,Giare:2025pzu}, and explore the interplay of our constraints with cosmological probes of dark energy interactions (see for instance Refs.~\cite{Vagnozzi:2019kvw,BeltranJimenez:2020iyx,BeltranJimenez:2021wbq,Ferlito:2022mok,BeltranJimenez:2024lml,Jimenez:2024lmm,BeltranJimenez:2025yad}). More generally, while all cosmological evidence for dark energy so far has been purely gravitational, our work shows that terrestrial experiments are well placed to probe dark energy's non-gravitational interactions and, conceivably, open the way towards direct detection of dark energy. We leave a detailed study of these interesting directions to follow-up work.

\begin{acknowledgments}
\noindent We thank the referee for their careful review of our manuscript and code. We thank Tom O'Shea and Jinqiang Ye for helpful discussions, and Clare Burrage for comments on this work. This publication is based upon work from the COST Action CA21106 ``COSMIC WISPers in the Dark Universe: Theory, astrophysics and experiments'' (Cosmic WISPers) and CA21136 ``Addressing observational tensions in cosmology with systematics and fundamental physics'' (CosmoVerse), both supported by COST (European Cooperation in Science and Technology). G.-W.Y.\ and S.V.\ acknowledge support from the University of Trento and the Provincia Autonoma di Trento (PAT, Autonomous Province of Trento) through the UniTrento Internal Call for Research 2023 grant ``Searching for Dark Energy off the beaten track'' (DARKTRACK, grant agreement no.\ E63C22000500003), and from the Istituto Nazionale di Fisica Nucleare (INFN) through the Commissione Scientifica Nazionale 4 (CSN4) Iniziativa Specifica ``Quantum Fields in Gravity, Cosmology and Black Holes'' (FLAG). A.C.D.\ acknowledges partial support from the Science and Technology Facilities Council (STFC) through STFC consolidated grant ST/T000694/1. M.G.\ acknowledges support from the Spanish Agencia Estatal de Investigación under grant PID2019-108122GB-C31, funded by MCIN/AEI/10.13039/501100011033, and from the ``European Union NextGenerationEU/PRTR'' (Planes complementarios, Programa de Astrof\'{i}sica y F\'{i}sica de Altas Energ\'{i}as). M.G.\ also acknowledges support from grant PGC2022-126078NB-C21, ``A\'{u}n m\'{a}s all\`{a} de los modelos est\'{a}ndar'', funded by MCIN/AEI/10.13039/501100011033 and ``ERDF A way of making Europe''. Additionally, M.G.\ acknowledges funding from the European Union's Horizon 2020 research and innovation programme under the European Research Council (ERC) grant agreement ERC-2017-AdG788781 (IAXO+). L.V.\ acknowledges support from the INFN CSN4 Iniziativa Specifica ``Quantum Universe'' (QGSKY), the National Natural Science Foundation of China (NSFC) through grant no.\ 12350610240 ``Astrophysical Axion Laboratories'', and the State Key Laboratory of Dark Matter Physics at the Shanghai Jiao Tong University. L.V.\ additionally thanks the Tsung-Dao Lee Institute and the Xplorer Symposia Organization Committee of the New Cornerstone Science Foundation for hospitality during the final stages of this work.
\end{acknowledgments}

\section*{Data Availability Statement}
\noindent The data and code supporting the results of this study are publicly available~\cite{chameleoncode}.

\appendix
\section{Solar model}
\label{app:solarmodel}

\begin{figure*}[!htbp]
\includegraphics[width=0.8\linewidth]{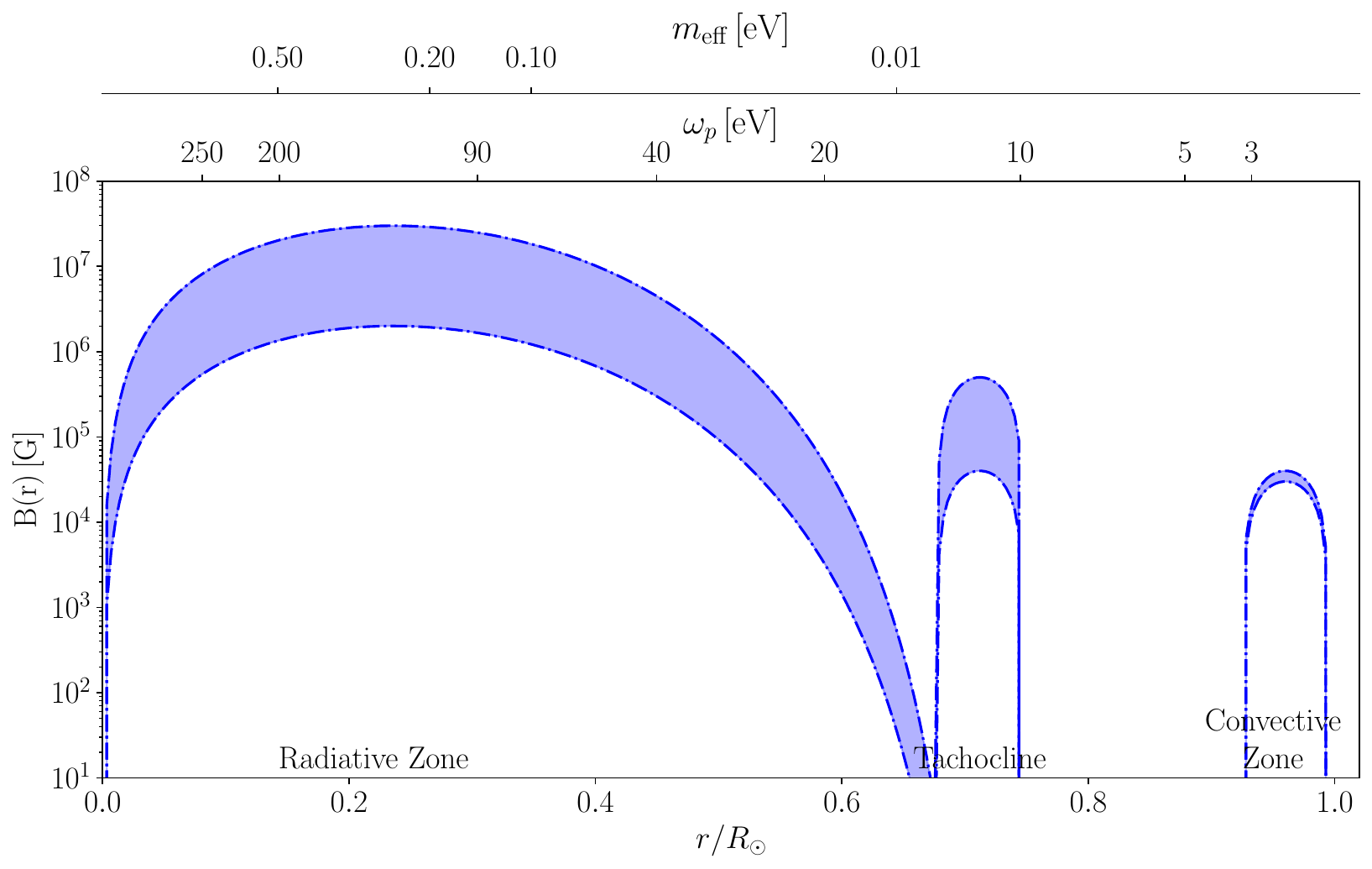}
\caption{Radial profile of the solar magnetic field $B(r)$, as a function of distance from the solar core, normalized by the solar radius $R_{\odot}$. We label the three regions of the solar interior, namely the radiative zone ($r \lesssim 0.7R_{\odot}$), the tachocline ($r \sim 0.7R_{\odot}$), and the convective zone ($r \gtrsim 0.7R_{\odot}$). Within each region, the width of the blue band reflects the uncertainties on the corresponding magnetic field strengths, i.e.\ $B_{\text{rad}} \in [2,30] \times 10^6\,{\text{G}}$ within the radiative zone, $B_{\text{tach}} \in [4,50] \times 10^4\,{\text{G}}$ within the tachocline, and $B_{\text{conv}} \in [3,4] \times 10^4\,{\text{G}}$ within the convective zone. The two upper x-axes display, respectively, the effective chameleon mass $m_{\text{eff}}$ (uppermost axis) and the plasma frequency $\omega_p$ (intermediate axis), both in ${\text{eV}}$. The effective chameleon mass is shown for the representative choice of chameleon parameter $\beta_m=10^2$, $\Lambda=2.4\,{\text{meV}}$, and $n=1$. }
\label{fig:bfield}
\end{figure*}

An important role in our analysis is played by the model for the solar interior, and several non-magnetic standard solar models are available in the literature (see e.g.~\cite{Grevesse:1998bj,Asplund:2009fu,Caffau:2010qc,Scott:2014lka,Scott:2014mka,vonSteiger:2016ghw,Vagnozzi:2016cmr,Vinyoles:2016djt,Vagnozzi:2016pjr,Asplund:2021ghw} for examples). In what follows, for consistency with Paper~II, we adopt the AGSS09 model~\cite{Asplund:2009fu}, where the Sun is modelled as a spherically symmetric and quasi-static star, whose evolution is described by a set of differential equations for its luminosity, radius, age and composition. While this is not the most recent model in the literature, we do not expect to observe significant changes were we to switch to a more recent model; moreover, the assumption of a quasi-static star only truly fails in the convective zone, whose contribution to the solar chameleon flux and resulting electron recoil signal is nevertheless completely negligible (as can be seen in Fig.~\ref{fig:signalcomponents}. In addition, large-scale magnetic fields, not explicitly included in standard solar models such as AGSS09, but inferred from helioseismic and dynamo studies, play a key role in our analysis.

Following Paper~II, we model the solar magnetic field as a quadrupolar configuration divided into three radial regions, each characterized by its own field strength and profile: from innermost to outermost, these are the radiative zone (``rad''), the tachocline (``tach''), and the convective zone (``conv''). In spherical coordinates $(r,\vartheta,\phi)$ centered around the solar core, the overall magnetic field profile we adopt is given as follows:
\begin{equation}
\mathbf{B}(r,\vartheta)=3B(r)\cos(\vartheta)\sin(\vartheta)\hat{\mathbf{e}}_\phi\,,
\label{eq:magneticfield}
\end{equation}
where the radial profile $B(r)$ is given by the following:
\begin{widetext}
\begin{equation}
B(r)=
\begin{cases}
B_{\text{rad}}(1+\lambda) \left (1+\dfrac{1}{\lambda} \right ) ^{\lambda} \left (\dfrac{r}{r_{\text{rad}}} \right ) ^2 \left [ 1- \left ( \dfrac{r}{r_{\rm rad}}\right)^2 \right ] ^{\lambda} & (r<r_{\text{rad}})\,, \quad \ \lambda\equiv 10r_{\text{rad}}+1\,; \\
B_{\text{tach}} \left [ 1-\left(\dfrac{r-r_{\text{tach}}}{d_{\text{tach}}} \right ) ^2 \right] & (\vert r-r_{\text{tach}} \vert <d_{\text{conv}})\,; \\
B_{\text{conv}} \left [ 1-\left(d\frac{r-r_{\text{conv}}}{d_{\text{conv}}} \right ) ^2 \right ]  & (\vert r-r_{\text{conv}} \vert <d_{\text{conv}}\,.\\
\end{cases}
\label{eq:bfield}
\end{equation}
\end{widetext}
The separation into three regions is clear from Eq.~(\ref{eq:bfield}): the radiative zone extends up to $r_{\text{rad}}=0.712R_{\odot}$, whereas the tachocline is centered at $r_{\text{tach}}=0.712R_{\odot}$ with a half-width of $d_{\text{tach}}=0.035R_{\odot}$, and the convective zone is centered at $r_{\text{conv}}=0.96R_{\odot}$ with a thickness of $d_{\text{conv}}=0.035R_{\odot}$. As for the magnetic field strength in the three zones, there is considerable uncertainty on the values of the three associated parameters $B_{\text{rad}}$, $B_{\text{tach}}$, and $B_{\text{conv}}$. Following Paper~II, we adopt the highly conservative ranges $B_{\text{rad}} \in [2,30] \times 10^6\,{\text{G}}$, $B_{\text{tach}} \in [4,50] \times 10^4\,{\text{G}}$, and $B_{\text{conv}} \in [3,4]\times 10^4\,{\text{G}}$. The radial profile of the magnetic field described by Eq.~(\ref{eq:bfield}) is shown in Fig.~\ref{fig:bfield}, with the width of the blue shaded bands reflecting the uncertainty in the strength of the magnetic field within each of the three zones. In the same figure, the two upper x-axes display, respectively, the effective chameleon mass $m_{\text{eff}}$ (uppermost axis) and the plasma frequency $\omega_p$ (intermediate axis), with the former computed assuming the benchmark values of $\beta_m=10^2$, $\Lambda=2.4\,{\text{meV}}$, and $n=1$. Throughout the work, we have assumed the mean value of the magnetic field strength when deriving predictions for the electron recoil event rate in XENONnT. Nevertheless, we note that the uncertainties on the magnetic field strength have a marginal impact on our results, given that the chameleon spectrum and corresponding event rate are completely dominated by Primakoff production in the electric fields of electrons and ions (as is very clear from Fig.~\ref{fig:signalcomponents}).

\section{Posterior distributions of the chameleon parameters}
\label{app:posteriors}

\begin{figure}[!htbp]
\centering
\includegraphics[width=0.9\linewidth]{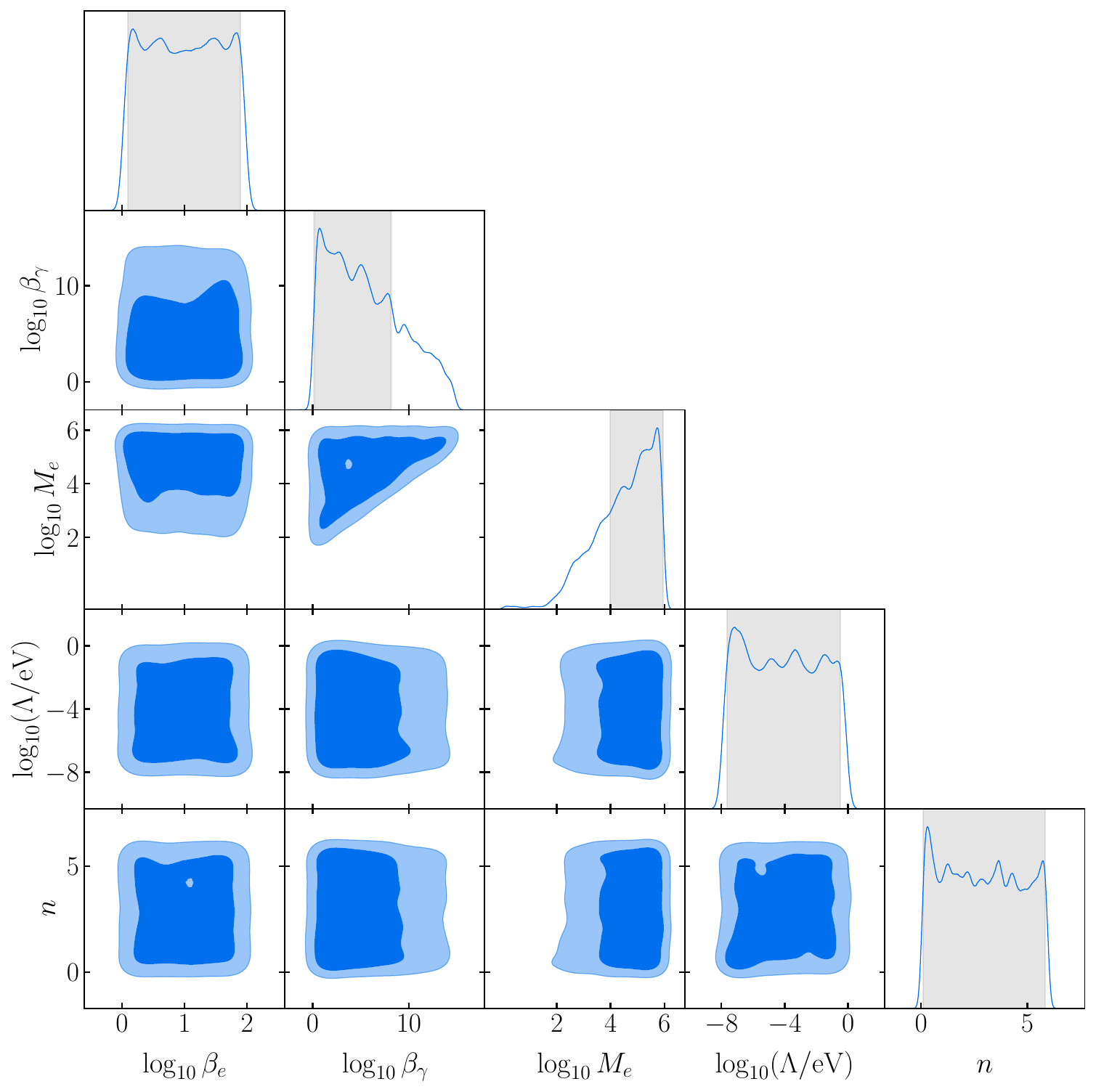}
\caption{Corner plot showing 2D joint and 1D marginalized posterior probability distributions for the chameleon parameters $\log_{10}{\beta_e}$, $\log_{10}{\beta_{\gamma}}$, $\log_{10}{M_e}$, $\log_{10}(\Lambda/{\text{eV}})$, and $n$, obtained from our exploratory Bayesian analysis of XENONnT electron recoil data, with shaded regions corresponding to the 68\% and 95\% credible intervals.}
\label{fig:posteriors}
\end{figure}

In Fig.~\ref{fig:posteriors} we show the full 2D joint and 1D marginalized posterior distributions for the chameleon parameters $\log_{10}{\beta_e}$, $\log_{10}{\beta_{\gamma}}$, $\log_{10}{M_e}$, $\log_{10}(\Lambda/{\text{eV}})$, and $n$, obtained from our exploratory Bayesian analysis of XENONnT electron recoil data. It is clear that $\beta_e$, $\Lambda$, and $n$ remain essentially unconstrained. As for $\beta_{\gamma}$ and $M_e$, we see a clear degeneracy/negative correlation between the two. This correlation is precisely along the direction described by the effective coupling strength $\beta_{\text{eff}}$, confirming that the fit to the XENONnT signal is mostly driven by this derived parameter. The posteriors given in Fig.~\ref{fig:posteriors} show that, although the full parameter space is in principle five-dimensional, it is mostly this specific combination of $\beta_{\gamma}$ and $M_e$ that matters for our analysis. This validates our earlier choice of presenting upper limits on $\beta_{\text{eff}}$ in Fig.~\ref{fig:betaeffupperlimit}, for which we recall that we find a 95\%~C.L. upper limit of $\log_{10}\beta_{\text{eff}}<-6.9$, as discussed in the main text.

\bibliography{solarchameleon}

\end{document}